\documentclass[fleqn, usenatbib]{mnras}

\usepackage{amsmath}
\usepackage{amsfonts}
\usepackage{amsbsy}
\usepackage{amssymb}
\usepackage{amsbsy}
\usepackage{makecell}
\usepackage[T1]{fontenc}
\usepackage[utf8]{inputenc}
\usepackage{ae,aecompl}
\usepackage{subfigure}
\usepackage{verbatim}
\usepackage{graphicx}
\usepackage{color}
\usepackage{multirow}
\usepackage{url}
\usepackage{lastpage}

\usepackage{acronym} 
\usepackage{mathtools} 
\usepackage{cleveref} 

\usepackage{caption}

\newacro{SI}{streaming instability}
\newcommand{\SI}{\ac{SI}}

\newacro{RDI}{resonant drag instability}
\newcommand{\RDI}{\ac{RDI}}
\newacroplural{RDI}{resonant drag instabilities}
\newcommand{\RDIs}{\acp{RDI}}

\newacro{PPD}{protoplanetary disc}
\newcommand{\PPD}{\ac{PPD}}
\newacroplural{PPD}{protoplanetary discs}
\newcommand{\PPDs}{\acp{PPD}}

\newacro{ODE}{ordinary differential equation}
\newcommand{\ODE}{\ac{ODE}}
\newacroplural{ODE}{ordinary differential equations}

\newacro{IVP}{initial value problem}

\newcommand\bb[1]{\mbox{\boldmath{$#1$}}}

\newcommand{\pD}[2]{\partial_{#1} #2} 
\newcommand{\D}[2]{{\rm d}_{#1} #2} 
\newcommand\grad{\bb{\nabla}} 

\newcommand\bcdot{\,\bb{\cdot}\,}

\newcommand{\St}{\text{St}} 
\newcommand{\dtg}{\mu} 

\newcommand{\ex}{\ensuremath{\bb{e}_{x}}} 
\newcommand{\ey}{\ensuremath{\bb{e}_{y}}} 
\newcommand{\ez}{\ensuremath{\bb{e}_{z}}} 

\newcommand{\bx}{\mathbf{x}} 
\newcommand{\bk}{\mathbf{k}} 
\newcommand{\hbk}{\hat{\mathbf{k}}} 
\newcommand{\hk}{\hat{k}} 
\newcommand{\bu}{\mathbf{u}} 

\newcommand{\e}{\mathrm{e}} 

\title[Explaining the streaming instability]{The physical mechanism of the streaming instability}
\author[N.~Magnan, T.~Heinemann \& H.~Latter]{
    Nathan Magnan$^{1}$\thanks{Contact e-mail: \href{mailto:nmtm2@cam.ac.uk}{nathan.magnan@maths.cam.ac.uk}}, 
    Tobias Heinemann$^{2}$, 
    and Henrik N.~Latter$^{1}$ 
    \\
    $^{1}$DAMTP, University of Cambridge, CMS, Wilberforce Road, Cambridge CB3 0WA, UK. \\
    $^{2}$Niels Bohr International Academy, Niels Bohr Institute, Blegdamsvej 17, 2100 Copenhagen, Denmark
    }

\begin{document}

\label{firstpage}
\pagerange{\pageref{firstpage}--\pageref{lastpage}}

\maketitle

\begin{abstract}
The main hurdle of planet formation theory is the metre-scale barrier. One of the most promising ways to overcome it is via the streaming instability (SI). Unfortunately, the mechanism responsible for the onset of this instability remains mysterious. It has recently been shown that the SI is a Resonant Drag Instability (RDI) involving inertial waves. We build on this insight and clarify the physical picture of how the SI develops, while bolstering this picture with transparent mathematics. Like all RDIs, the SI is built on a feedback loop: in the \lq forward action', an inertial wave concentrates dust into clumps; in the \lq backward reaction', those drifting dust clumps excite an inertial wave. Each process breaks into two mechanisms, a fast one and a slow one. At resonance, each forward mechanism can couple with a backward mechanism to close a feedback loop. Unfortunately, the fast-fast loop is stable, so the SI uses the fast-slow and slow-fast loops. Despite this last layer of complexity, we hope that our explanation will help understand how the SI works, in which conditions it can grow, how it manifests itself, and how it saturates.
\end{abstract}


\defcitealias{YoudinGoodman05}{YG05}
\defcitealias{SquireHopkins18a}{SH18a}
\defcitealias{SquireHopkins18b}{SH18b}
\defcitealias{SquireHopkins20}{SH20}
\defcitealias{Magnan24}{M24}

\section{Introduction}
\label{sec:intro}

A major problem in planet formation theory is the metre-scale barrier: we do not understand how pebbles assemble into planetesimals. Collisions between pebbles lead to bouncing and fragmentation rather than sticking \citep{Weidenschilling84}, and for small planetesimals, accretion become slower than the radial drift toward the star \citep{Weidenschilling77}.

One promising way to bridge this gap involves a mechanism called the \SI. This hydrodynamic instability afflicts gas and dust mixtures in proto-planetary discs (\citealt{YoudinGoodman05}, hereafter \citetalias{YoudinGoodman05}). The interesting thing is that it saturates by forming dust filaments and clumps \citep{JohansenYoudin07}. The clumps can be dense enough to collapse gravitationally and form planetesimals \citep{Johansen+07}. That being said, there is some debate about whether pre-existing turbulence \citep{Umurhan+20, ChenLin20, Schafer+20, SchaferJohansen22, XuBai22, Lim+24}, or a distribution of particle sizes~\citep{Krapp19, McNally21, ZhuYang21, YangZhu21} quench the~instability.

Here we address another difficulty, namely that the physics of the \SI\ is still unclear. This is problematic because it is hard to ask the right questions about a phenomenon we do not understand. It also makes it challenging to interpret numerical simulations. This has already motivated several attempts to try and explain the instability \citep{YoudinJohansen07, Jacquet+11, LinYoudin17}.

\enlargethispage{+0.3 \baselineskip}
Squire \& Hopkins recently made progress on that front. They discovered in \citealt{SquireHopkins18a} (hereafter \citetalias{SquireHopkins18a}) that the \SI\ is part of a larger class of instabilities called \RDIs. Those arise when the dust drifts relative to the gas and the drift velocity matches the phase velocity of one of the gas waves. Then in \citealt{SquireHopkins20} (hereafter \citetalias{SquireHopkins20}) they provided a physical picture for the \SI. The present paper is an effort to support and refine this picture with rigorous mathematics. Far from complicating things, we find that the algebra actually brings clarity. We hope that the steps of the derivation will guide our reader's intuition, thus making it less laborious to understand the~\SI.

In a previous paper, we explained the simplest of all \RDIs, the one built on sound waves (\citealt{Magnan24}, hereafter \citetalias{Magnan24}). We developed a mathematical framework to study the instability's eigenvectors. It showed that the \RDI\ is based on a positive feedback loop between two processes. In the first one, the converging flows of a sound wave concentrate dust. In the second one, a dust density wave excites or amplifies sound waves. A similar loop is at the heart of every \RDI, what varies is the gas wave and how it concentrates dust.

In this second paper, we apply the same framework to explain the more complicated but astrophysically relevant \SI. It is an \RDI\ built on inertial waves. Those are incompressive, so it is unclear how they can concentrate dust (a fundamentally compressive deed). Resolving this paradox is one of our key objectives. We suspect that our solution applies to all~inertial-wave \RDIs, but we focus on the particular case of~the \SI.

\enlargethispage{+0.3 \baselineskip}
The structure of the paper is as follows. We start in \S\ref{sec:intuition} with a summary of the physics of the \SI, which we hope will help the reader form an intuition for what is happening before we dive into the mathematics. Then in \S\ref{sec:governing_equations} we present the physical system and its governing equations, and in \S\ref{sec:core} we explain how the instability develops. Finally, we conclude in \S\ref{sec:conclusion}.

\newcommand{\bv}{\mathbf{v}} 

\begin{figure*}
    \begin{minipage}{0.49 \linewidth}
        \centering
        \vspace*{-0.65 \baselineskip}
        \includegraphics[width = \linewidth]{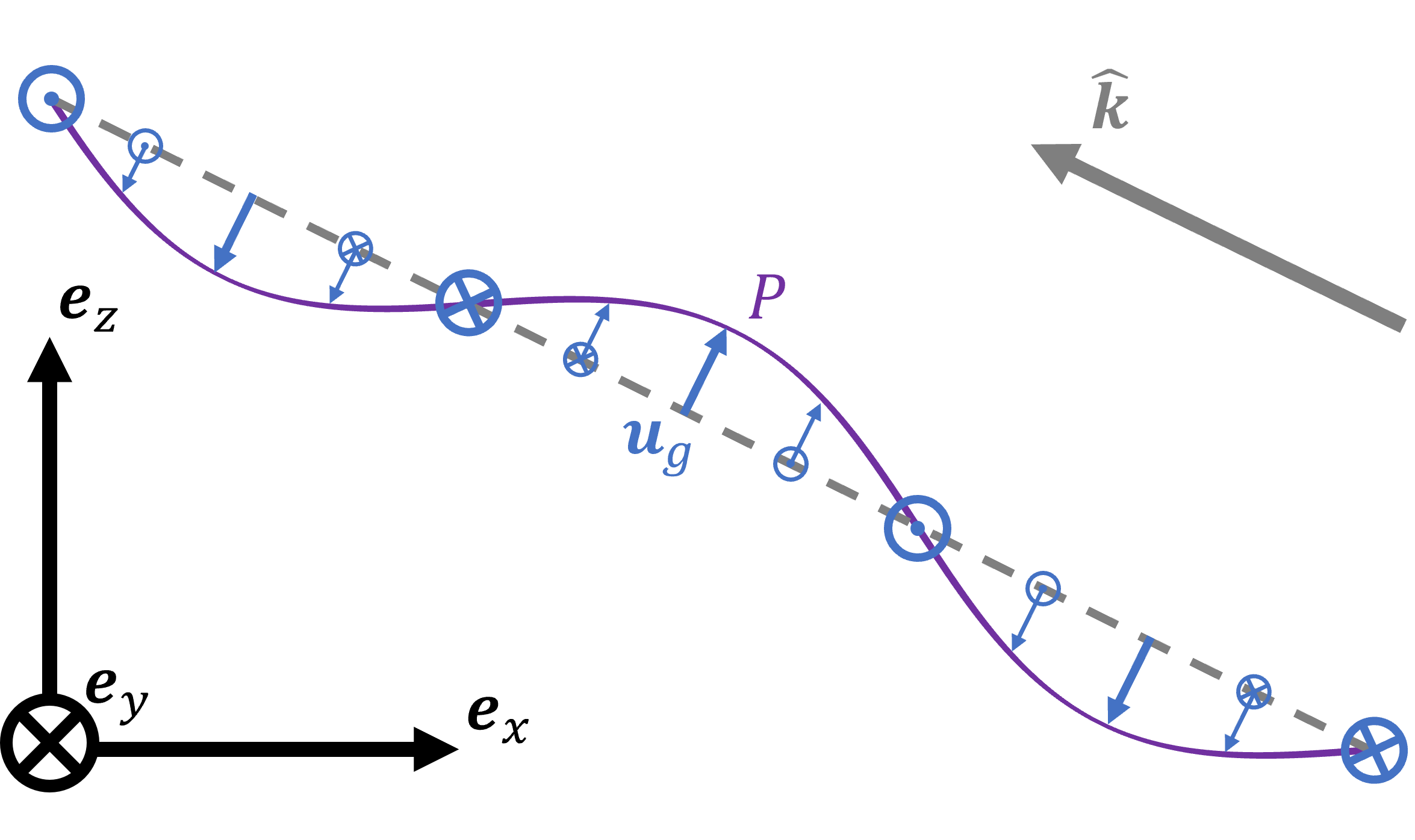}
        \vspace*{1.5 \baselineskip}
        \caption{
            Schematic drawing of an inertial wave. In black is the coordinate system, with $\ex$ giving the radial direction, $\ey$ the azimuthal, and $\ez$ the vertical. $\ey$ is perpendicular to the drawing plane and pointing away from the reader. Blue denotes the gas' velocity perturbation, and the purple line represents the pressure perturbation. Finally, the grey arrow represents the wavevector $\bk$. Note that the wave is elliptically polarised and transverse.
        }
        \label{fig:inertial_wave}
        \vspace*{-1 \baselineskip}
    \end{minipage}
    \hfill
    \begin{minipage}{0.49 \linewidth}
        \centering
        \vspace{-1.5 \baselineskip}
        \includegraphics[width = 0.7 \linewidth]{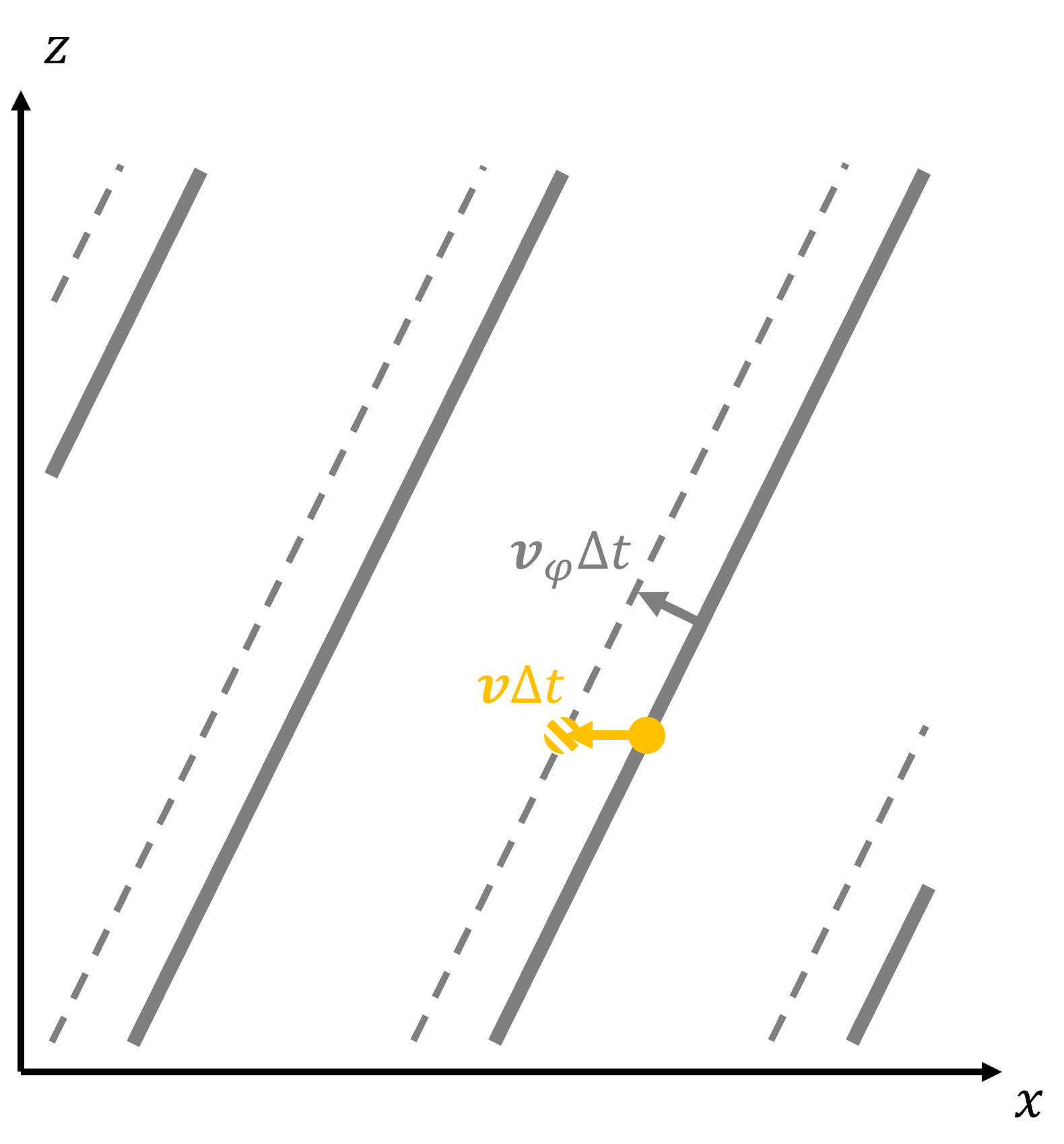}
        \vspace*{- 0.5 \baselineskip}
        \caption{
            Schematic drawing of the resonance condition (\textit{cf.} \S\ref{sub:intuition_setup}). The solid grey lines represent the crests of an inertial wave at time~${ t }$, and the solid orange dot is a dust particle that sits on a wave crest. The dashed grey lines and the dashed orange circle represent the situation at a time $\Delta t$ later. $ \mathbf{v}_{\varphi} $ is the inertial wave's phase speed and $\bv$ is the dust's drift speed. If the resonance condition holds, the dust particle remains on the same wave crest at all times.
        }
        \vspace*{-1 \baselineskip}
        \label{fig:resonance_condition_drawing}
    \end{minipage}
\end{figure*}

\section{Outline of the mechanism}
\label{sec:intuition}

In this first section, we give a quick summary of how the streaming instability develops. We do not use any algebra or calculus yet. Our goal here is rather to give the reader an intuition for how perturbations grow. This will help clarify and motivate the mathematics of \S\ref{sec:core}.

\vspace{-0.8 \baselineskip}
\subsection{The setup: inertial waves, radial drift and resonance}
\label{sub:intuition_setup}

Before we begin, let us define our setup. We consider a small patch of a \PPD\ modelled using the shearing box \citep{GoldreichLyndenBell65, Hawley+95, LatterPapaloizou17}. This box is filled with incompressible gas, plus a small amount of dust modelled as a pressure-less fluid. The two components act upon each other via Stokes or Epstein drag.

The disc's stable radial angular momentum stratification allows the gas to support inertial waves. These are transverse waves restored by the Coriolis force (Fig.~\ref{fig:inertial_wave}). Because of the \PPD's shear, only axisymmetric inertial waves can propagate without being damped. Their frequencies are ${ \pm \hk_{z} \kappa }$, with $\bk$ the wave-vector and $\kappa$ the disc's local epicyclic frequency. For more details on inertial waves, see \cite{Balbus03}.

In a \PPD, pressure generally decreases with radius. This radial pressure gradient indirectly causes the dust to drift towards the star \citep{Weidenschilling77}. Precisely, the pressure gradient supports the gas radially, making it sub-Keplerian. Conversely, the dust ignores pressure, so it remains Keplerian and experiences a headwind. Drag induces angular momentum exchanges between dust and gas, so the dust loses momentum. Consequently, it drifts inwards (and azimuthaly).

For any drift velocity~$\bv$, there exists an inertial wave that propagates at just the right phase velocity $ \mathbf{v}_{\varphi} $ to keep up with the dust. Specifically, we demand that ${ \bv \bcdot \bk = \mathbf{v}_{\varphi} \bcdot \bk }$ where~$\bk$ is the wave's vector. Figure~\ref{fig:resonance_condition_drawing} shows how this condition ensures that a dust particle sitting on a wave crest remains on that crest. For reasons that will become clear later, we shall say that the dust is \lq in resonance' with the inertial wave. The goal of the present paper is to study this resonance.

\subsection{Forcing of a dust density perturbation by an inertial wave}
\label{sub:intuition_forward_action}

First consider the dust as passive tracers (\textit{i.e.} test particles). The gas behaves as in isolation, so inertial waves propagate freely. The resonant inertial wave can elicit a strong response from the dust via two mechanisms. 

\begin{figure*}
    \vspace{-1 \baselineskip}
    \centering
    \includegraphics[width = 1 \linewidth]{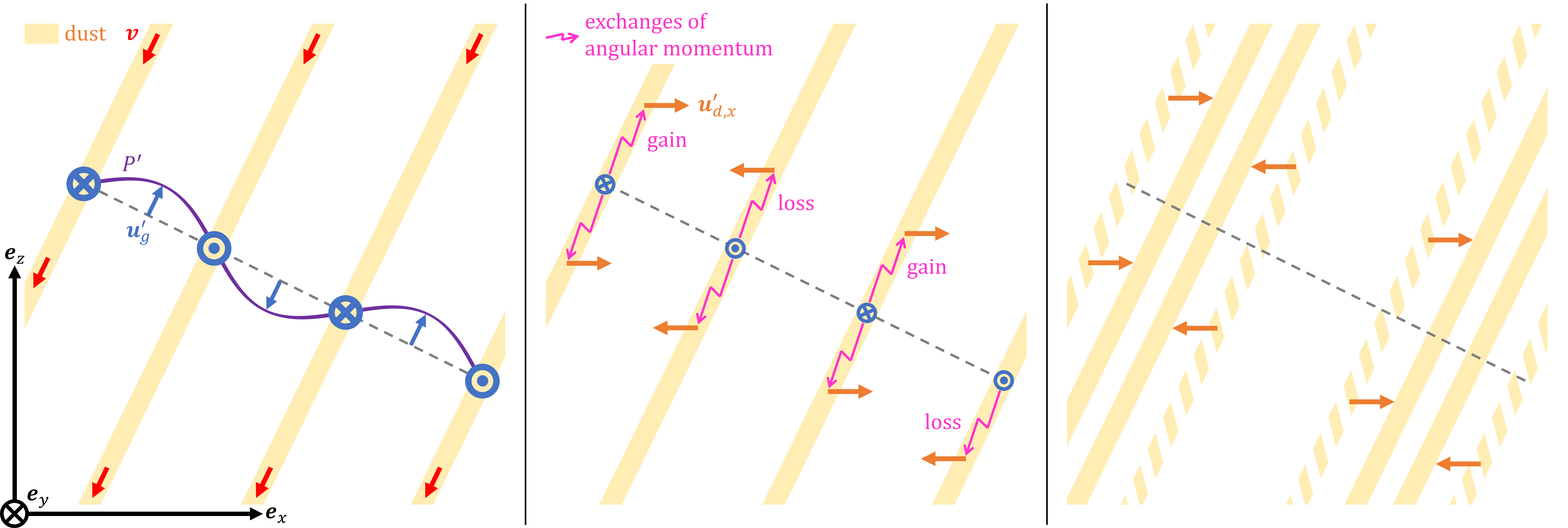}
    \vspace*{- 1.5 \baselineskip}
    \caption{
        Comic strip explaining the fast pressure-bump mechanism of the forward action (\textit{cf.} \S\ref{sub:intuition_forward_action}).
        \textit{Left:} Imagine an inertial wave propagating freely. Blue denotes the gas velocity perturbation, and purple the pressure perturbation. Now imagine there is also some dust, and assume resonance. If we draw things in the reference frame that travels with the inertial wave, then the background dust drift~(red arrows) is contained within waveplanes. Consequently, we need to understand how entire \lq planes' of dust (orange) are displaced by the inertial wave.
        \textit{Centre:} The gas' azimuthal velocity perturbations mean that in some places, the dust experiences more headwind than usual, so it loses angular momentum faster, and drifts towards the star faster. Conversely, there are places where the dust's radial drift slows down. This is equivalent to radial dust velocity perturbations (orange arrows). 
        \textit{Right:} Those dust motions are compressive. Note that the regions of dust accumulation are pressure maxima.
    }
    \label{fig:Forward_action_fast}
\end{figure*}

The strongest one relies on the well-known propensity of pressure bumps to trap dust particles. It is sketched out in Fig.~\ref{fig:Forward_action_fast}, in the reference frame that moves with the resonant inertial wave. This wave induces variations in the gas' azimuthal speed $\bu_{g, y}$ (in blue). This means that a dust particle located on a wave crest in $\bu_{g, y}$ feels less headwind than usual, while a particle located on a wave trough feels more headwind than usual. Consequently, the dust's rate of angular momentum loss and its radial drift speed are perturbed. The latter is equivalent to radial dust velocity perturbations (in brown), which compress the dust in pressure maxima.

Crucially, since we are at resonance, the dust's radial drift lets particles remain on pressure crests. So, the dust density increases indefinitely as more and more dust arrives. This results in an algebraic instability we call the \lq forward action'.

\begin{figure*}
    \vspace{-1 \baselineskip}
    \centering
    \includegraphics[width = 1 \linewidth]{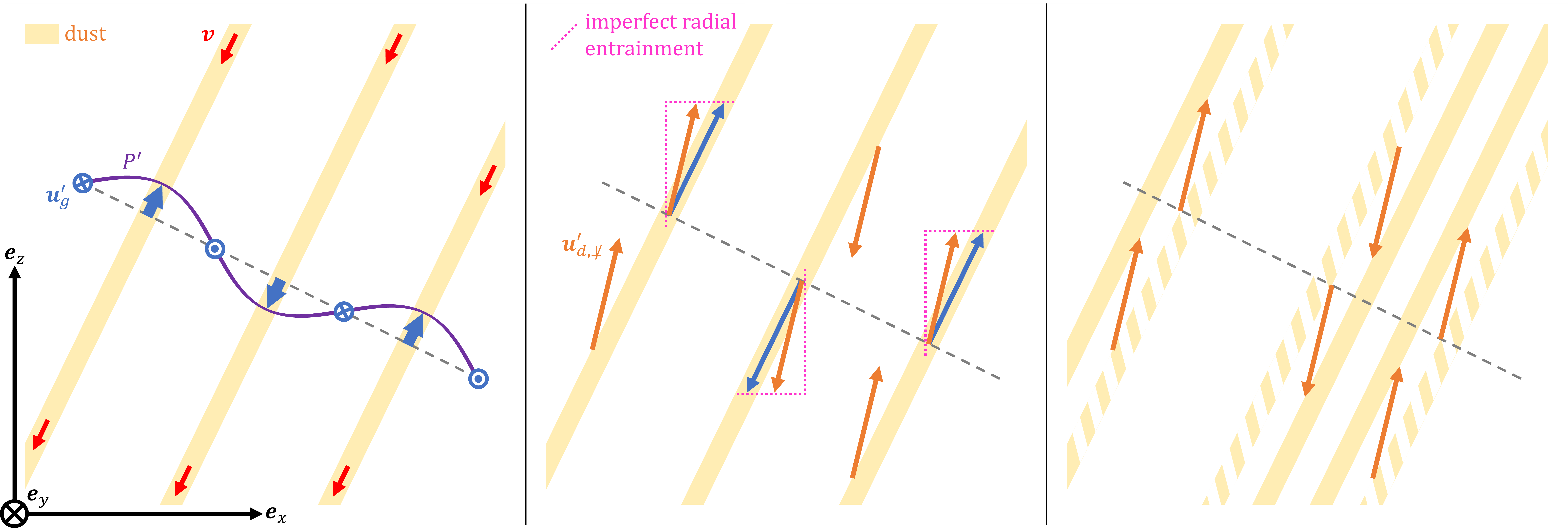}
    \vspace*{- 1.5 \baselineskip}
    \caption{
        Comic strip explaining the slow radial-wind mechanism of the forward action (\textit{cf.} \S\ref{sub:intuition_forward_action}).
        \textit{Left:} Same as the left panel of Fig.~\ref{fig:Forward_action_fast}, except that we consider different dust planes.
        \textit{Centre:} The gas is incompressible, so its radial-vertical velocity perturbations are contained within waveplanes. Those winds entrain the dust, resulting in radial-vertical dust velocity perturbations (orange arrows). Crucially, dust follows the gas perfectly in the vertical direction but not in the radial direction, due to the mechanism of \S\ref{sec:dust_grain_pusehd_radially}.
        \textit{Right:} The dust has a component of motion perpendicular to the waveplanes, and that causes compression. This time, dust accumulates in pressure nodes.
    }
    \label{fig:Forward_action_slow}
\end{figure*}

The second mechanism, sketched in Fig.~\ref{fig:Forward_action_slow}, is weaker. The inertial wave induces gas motions in the radial-vertical plane. These winds try to entrain the dust, but orbital mechanics tell us that a particle pushed outwards does not experience any long-term gain in radius, it only oscillates. Indeed, a radial push does not provide any angular momentum, so the particle cannot stay on an outer orbit. Of course, in a \PPD, there is gas in near-Keplerian rotation, so the displaced dust particle can receive angular momentum and follow the radial wind. But still, the orbital mechanics effect described above makes the dust's radial motions slower than the gas'. Now since the dust moves vertically as quickly as the gas, and since the gas' radial-vertical motions are fine-tuned to be perpendicular to $\bk$, the dust must have a small component of motion parallel to $\bk$. This results in dust convergence towards pressure nodes. For more details on this \lq imperfect radial entrainment' process, see \S\ref{sec:dust_grain_pusehd_radially}.

\subsection{Forcing of an inertial wave by a dust density perturbation}
\label{sub:intuition_backward_reaction}

To each action, there is an equal and opposite reaction. So the dust must also exert a drag force on the gas, we call it the back-reaction. Through this force, the gas can react strongly to a dust density perturbation -- provided we are at resonance.

\begin{figure*}
    \centering
    \includegraphics[width = 1 \linewidth]{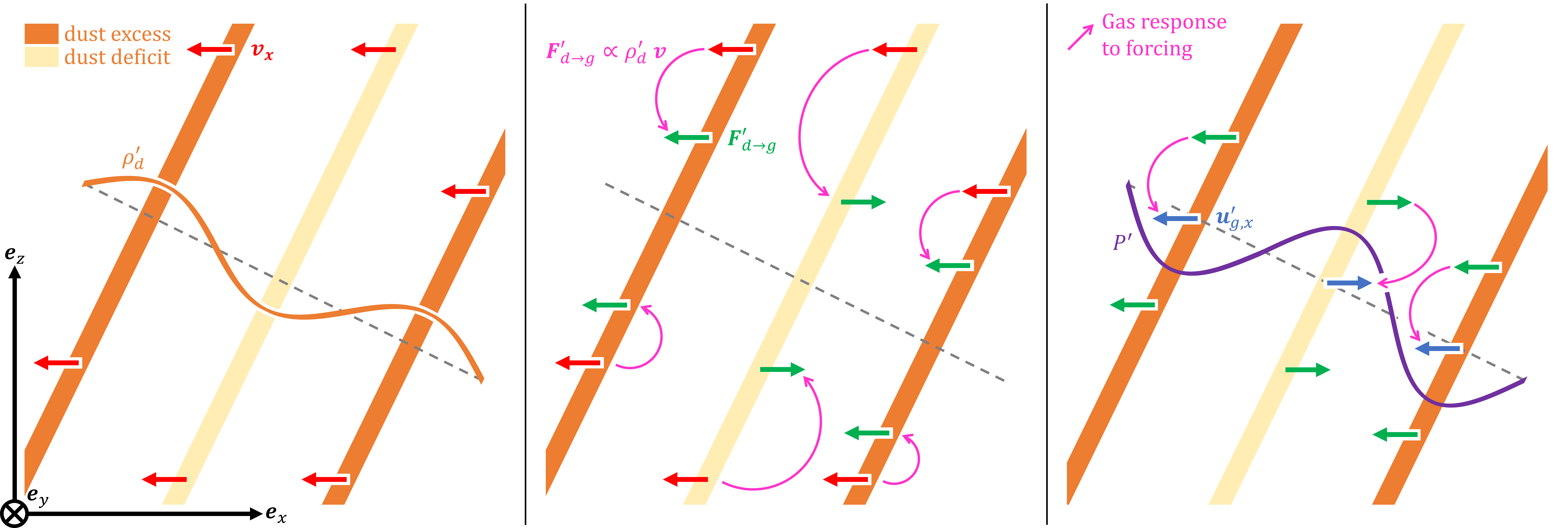}
    \vspace*{- 1.5 \baselineskip}
    \caption{
        Comic strip explaining the fast radial mechanism of the backward reaction (\textit{cf.} \S\ref{sub:intuition_backward_reaction}).
        \textit{Left:} Imagine a dust density wave (orange line) drifting freely, unaffected by the gas. We use two shades of orange to indicate waveplanes that contain more or less dust than at equilibrium. Contrary to Figs.~\ref{fig:Forward_action_fast}~and~\ref{fig:Forward_action_slow}, we draw things in the standard reference frame, so the background dust drift (red arrows) is mostly radial.
        \textit{Centre:} As they drift, dust grains entrain the gas around them. This force of the dust on the gas is called the back-reaction. It is already present at equilibrium, but when the dust density wave causes a dust density anomaly, it also causes a back-reaction anomaly~\smash{${ F_{d\rightarrow g}^{'} }$} (green arrows) proportional to the dust density perturbation ${ \rho_{d}^{'} }$ and to the dust drift $\bv$.
        \textit{Right:} Through this force, the radial component of the dust drift causes radial gas velocity perturbations (blue arrows). Those gas movements are exactly those of the inertial wave whose pressure is in phase opposition with the dust density. And indeed, the gas responds to the forcing by amplifying that wave.
    }
    \label{fig:Backward_reaction_fast}
\end{figure*}

To see how this works, consider a dust density wave, \textit{i.e.} a sinusoidal perturbation of the dust density advected by the radial drift. Note that the dust's velocity is unperturbed, and that the wave's vector $\bk$ may well have a vertical component. In regions where there is more dust than at equilibrium, the back-reaction is stronger than at equilibrium. Figure~\ref{fig:Backward_reaction_fast} shows that the radial component of this perturbed back-reaction has the right spatial profile to accelerate the radial flows of an inertial wave, thereby amplifying that wave.

\mbox{Since \!\! we \!\! are \!\! at \!\! resonance, \!\!\! the \!\!\! dust \!\!\! perturbation \!\!\! drifts \!\!\! with} \mbox{the \!\!\! inertial \!\!\! wave, \!\!\! so \!\!\! the \!\!\! back-reaction maintains \!\!\! the \!\!\! right pro}-\mbox{file \!\! to \!\! continually \!\! strengthen \!\! the \!\! wave. \!\!\! This \!\!\! leads \!\!\! to \!\!\! a \!\!\! second} \mbox{algebraic instability, which we call the \lq backward reaction'.}

\begin{figure*}
    \centering
    \includegraphics[width = 1 \linewidth]{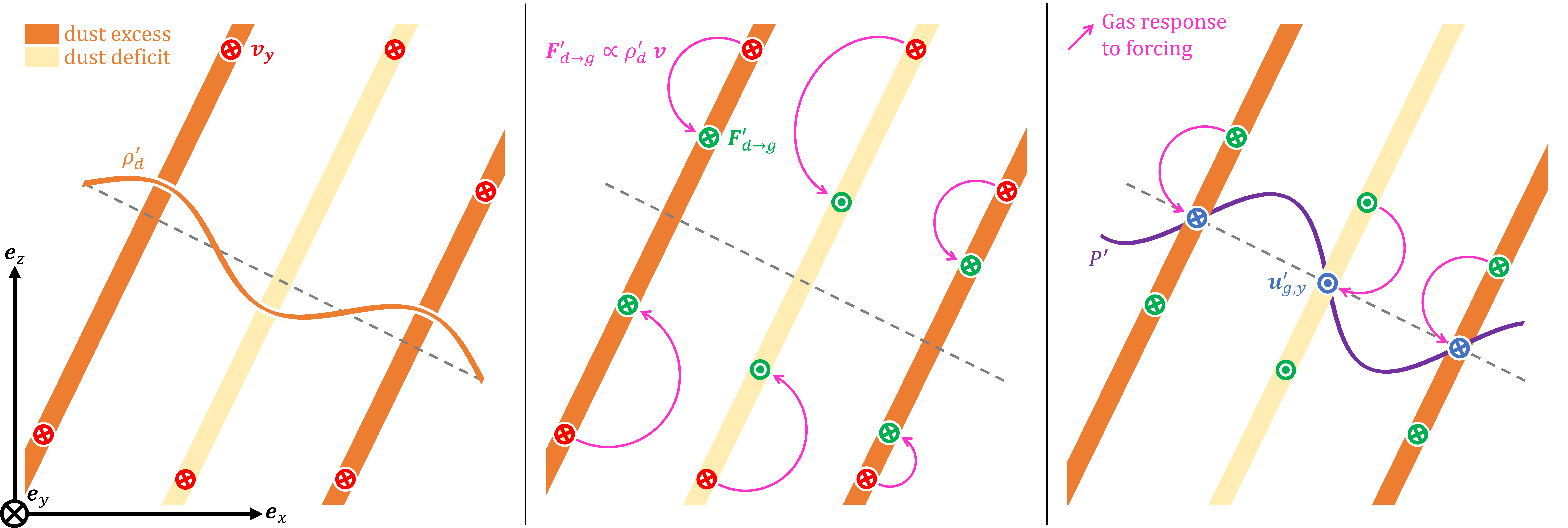}
    \vspace*{- 1.5 \baselineskip}
    \caption{
        Same as Fig.~\ref{fig:Backward_reaction_fast}, except that we focus on the azimuthal component of the dust drift. It drives the slow azimuthal mechanism of the backward reaction (\textit{cf.} \S\ref{sub:intuition_backward_reaction}), which amplifies the inertial wave whose pressure is dephased by ${ \pi / 2 }$ with respect to the dust density.
    }
    \label{fig:Backward_reaction_slow}
\end{figure*}

The background dust drift also has a small azimuthal component $v_{y}$, and so does the back-reaction force. Figure~\ref{fig:Backward_reaction_slow} shows that this azimuthal component of the perturbed back-reaction has just the right spatial profile to amplify the inertial wave whose pressure is dephased by $ { \pi / 2 }$ with respect to the dust density. This azimuthal mechanism is much slower than the radial mechanism of Fig.~\ref{fig:Backward_reaction_fast}.

\vspace{-1 \baselineskip}
\subsection{Combining these two processes in a feedback loop}
\label{sub:intuition_combine} 

So far, we have seen two isolated processes. In the forward action, an inertial wave creates or amplifies dust density waves. In the backward reaction, a dust density wave creates or amplifies inertial waves. We can easily imagine the two processes connect in a feedback loop where the inertial wave concentrates dust, and the ensuing dust density wave strengthens the inertial wave, which can then concentrate dust faster, \textit{etc}. This leads to an exponential instability, the \SI.

There is, however, one last layer of complexity. The forward action breaks up into the fast pressure-bump mechanism and the slow radial wind mechanism. Similarly, the backward reaction breaks into the fast radial mechanism and the slow azimuthal mechanism. Therefore, we could imagine four feedback loops: fast-fast, fast-slow, slow-fast, and slow-slow. 

It turns out that the fast-fast feedback loop is stable. Indeed, the pressure-bump mechanism concentrates dust in pressure maxima (Fig.~\ref{fig:Forward_action_fast}), whereas the radial backward mechanism amplifies the inertial wave whose pressure is anti-correlated with dust density (Fig.~\ref{fig:Backward_reaction_fast}). Because of this phase opposition, the two fast mechanisms work against each other.

\mbox{As a result, \!\! the \!\! \SI\ \!\!\! relies \!\!\! on \!\!\! the \!\!\! fast-slow \!\!\! and \!\!\! slow-fast \!\!\! loops.} Those two loops are of equal strength, and cooperate well enough that the instability grows faster than either loop.

\section{Presentation of the physical system}
\label{sec:governing_equations}

From this point onward, we shall detail and justify all the claims of \S\ref{sec:intuition}. We hope that mathematics will help some readers understand the physical picture described in this paper. 

Let us start with the setup (\S\ref{sub:intuition_setup}). Our physical system is equivalent to that studied by \citetalias{YoudinGoodman05}, and similar to but simpler than that studied by \citetalias{SquireHopkins18a}.

\subsection{Governing equations}
\label{sub:governing_equations}

\newcommand{\rhog}{\rho_{g}} 
\newcommand{\bug}{\ensuremath{\bb{u}_{g}}} 
\newcommand{\rhod}{\rho_{d}} 
\newcommand{\bud}{\ensuremath{\bb{u}_{d}}} 

We consider a gas and dust mixture, which we model as a two-fluid system. The gas is assumed to be incompressible, and is described by its density ${ \rhog }$, its speed ${ \bug }$, and its pressure ${ P }$. The dust is assumed to be pressure-less, and is described by its density ${ \rhod }$ and velocity ${ \bud }$. The two fluids are coupled by a linear drag force from the gas onto the dust, and its back-reaction from the dust onto the gas.

\newcommand{\Omegao}{\Omega_{0}} 

Typically, when studying a small-scale instability, one accounts for the fact that they are in a \PPD\ by using the shearing box \citep{GoldreichLyndenBell65, Hawley+95, LatterPapaloizou17}. This local model follows a reference point as it orbits with radius $r_{0}$ and frequency $\Omegao$ around the star. The local fluid dynamics are represented by Cartesian axes ${ (x, \, y, \, z) }$ pointing respectively in the radial, azimuthal, and vertical directions.

\newcommand{\Po}{\overline{P}} 

However, to study the \SI, we must add a non-standard ingredient to our box. Indeed, the \SI\ is powered by the radial drift of dust particles towards the star, which is itself powered by the disc's global pressure gradient. So, we need a radial background pressure gradient ${ \pD r \Po }$ in our shearing box.

\newcommand{\Omegag}{\Omega_{\text{g}}} 
\newcommand{\Omegad}{\Omega_{\text{d}}} 
\newcommand{\OmegaK}{\Omega_{\text{K}}} 

\newcommand{\Sg}{S_{\text{g}}} 
\newcommand{\Sd}{S_{\text{d}}} 

\newcommand{\fg}{f_{g}} 
\renewcommand{\fd}{f_{d}} 

\newcommand{\Omegago}{\Omega_{\text{g}, 0}} 

This introduces three complications. Firstly, the gas is pressure-supported, so its orbital frequency $\Omegag$ is slightly sub-Keplerian. Specifically, \smash{${ r \Omegag^{2} = r \OmegaK^{2} + \rhog^{-1} \pD r \Po }$}, where $\OmegaK$ is the Keplerian orbital frequency. Secondly, the dust orbits faster than the gas. This means that we need to decide which species our shearing box follows. We choose to follow the gas, so ${ \Omegao = \Omegago }$. It also means that the two fluids have different shear rates, $\Sg$ and $\Sd$. Note that $\Sg$ and $\Sd$ are defined \textit{in the absence of aerodynamical friction}. Finally, if we go back to the global cylindrical coordinates, we see that the radial geometric term in ${\bud \bcdot \grad \bud }$ \!is non-zero. Consequently, it must also appear in our Cartesian box’s dust momentum equation.

The most convenient way to account for these complications is to replace the tidal potential \smash{${ \phi_{t} }$} -- which only accounts for gravitational and centrifugal forces -- by two `effective' potentials \smash{${ \phi_{e}^{g} }$} and \smash{${ \phi_{e}^{d} }$}, one for each species. Specifically:

\newcommand{\Omegado}{\Omega_{\text{d}, 0}} 
\newcommand{\OmegaKo}{\Omega_{\text{K}, 0}} 

\vspace*{+0.5 \baselineskip}
\noindent $\bullet$ \textit{For the gas:} Since the shearing box follows the gas and since the gas is sub-Keplerian, the tidal potential takes the non-standard form 
\,\smash{${ \phi_{t} (\bx) \, = \, (\OmegaKo^{2} - \Omegago^{2}) \, r_{0} \, x - \Omegago \Sg \, x^{2} }$}. We group ${ \phi_{t} }$ with the background pressure $\Po$ so that the gas potential recovers its standard form, 
\begin{equation}
    \nonumber
    \phi_{e}^{g} = \phi_{t} + \frac{\pD r \Po |_{r_{0}}}{\rhog} \, x = - \, \Omegago \Sg \, x^{2}
\end{equation}

\noindent $\bullet$ \textit{For the dust:} The dust's effective potential, however, groups the sub-Keplerian tidal potential with the geometric term:
\begin{equation}
    \nonumber
    \phi_{e}^{d} = 2 \Omegago (\OmegaKo - \Omegago) \, r_{0} \, x - \Omegago \Sd \, x^{2} .
\end{equation}  
The two terms have the same size when ${ x \sim h^{2} r_{0} }$, where $h$ is the disc's local aspect ratio. Since this is the wavelength of the \SI's fastest-growing mode (\citetalias{YoudinGoodman05}), we must keep both terms. 
However, the exact values of the shear rates are non-essential, so we set ${ \Sd \! = \! \Sg \! = \! \frac{3}{2} \Omegago \! = S }$. We also drop the subscript $_{\text{g}, 0}$.

\enlargethispage{+1 \baselineskip}
\vspace{+0.5 \baselineskip}
\noindent All those assumptions are summed up by the equations
\begin{subequations}
    \label{eq:Navier-Stokes_SI}
    \begin{align}
        \grad \! \bcdot \bug \! &= 0 , \label{eq:Navier-Stokes_continuity_gas_SI} \\
        \! \pD t \bug \! + \! \bug \bcdot \! \grad \bug \! &= \! - \frac{\grad \! P'}{\rhog} \! - \! 2 \Omega \ez \! \wedge \! \bug \! - \!\! \grad \phi_{e}^{g} \! + \! \frac{\dtg}{\tau} (\bud \! - \! \bug) , \!\! \label{eq:Navier-Stokes_momentum_gas_SI} \\
        \pD t \rhod \! + \! \bud \bcdot \! \grad \rhod \! &= \! - \rhod \grad \! \bcdot \bud , \label{eq:Navier-Stokes_continuity_dust_SI} \\
        \! \pD t \bud \! + \! \bud \bcdot \! \grad \bud \! &= \! - 2 \Omega \ez \! \wedge \! \bud \! - \!\! \grad \phi_{e}^{d} \! - \! \frac{1}{\tau} (\bud \! - \! \bug) , \label{eq:Navier-Stokes_momentum_dust_SI}
    \end{align}
\end{subequations}
where $P'$ is the pressure perturbation relative to $\Po$,~${ \dtg = \rhod / \rhog }$ is the dust-to-gas ratio, ${ \tau }$ is the dust's stopping time, and $\ez$ is the vertical unit vector.

This physical system is equivalent to that of \citetalias{YoudinGoodman05}. It is also almost the same as that of \citetalias{SquireHopkins18a}, but slightly simpler. Namely, we assume that the dust's stopping time is fixed rather than variable. Since the gas is incompressible, this assumption is valid in the Stokes and Epstein drag regimes.

\subsection{Background equilibrium flow}
\label{sub:background_flow}

Let us now describe the equilibrium flow upon which infinitesimal perturbations will grow and cause the instability. We shall denote background variables with an overline ${ \overline{\phantom{a}} }$.

\newcommand{\bugo}{\ensuremath{\overline{\bb{u}}}_{g}} 
\newcommand{\budo}{\ensuremath{\overline{\bb{u}}}_{d}} 
\newcommand{\rhodo}{\overline{\rho_{d}}} 
\newcommand{\dtgo}{\overline{\dtg}} 

\newcommand{\fgo}{\overline{f_{g}}} 
\newcommand{\fdo}{\overline{f_{d}}} 

\newcommand{\bS}{\mathbf{S}} 
\newcommand{\bw}{\mathbf{w}} 
\renewcommand{\bv}{\mathbf{v}} 

To stay as close as possible to \citetalias{YoudinGoodman05}, we set
\begin{equation}
    \label{eq:background_SI_pressure}
    \pD r \Po = - 2 \, \eta \rhog r_{0} \Omega^{2} ,
\end{equation}
where ${ \eta \propto h^{2} }$ is a dimensionless number indicating the amount of pressure
support. This leads to
\begin{subequations}
    \label{eq:background_SI}
    \begin{align}
        \bugo &= \bS \bx - \fdo \, \bw , \label{eq:background_SI_gas} \\
        \budo &= \bS \bx + \fgo \, \bw + V \ey , \label{eq:background_SI_dust}
    \end{align}
\end{subequations}
where \smash{${ \bS = - S \, \ey \! \otimes \! \ex }$} is a matrix representing the shear, ${ \fd = 1 - \fg }$ is the dust mass fraction, ${ V \! = \! \eta \Omega r_{0} }$ is a scalar representing the dust-to-gas drift in the absence of coupling,
\!\!\begin{equation}
    \nonumber
    \bw = \frac{- 2 \fgo \St }{1 + (\fgo \St)^{2}} \left[ \ex + \frac{1}{2 \fgo \St} \, \ey \right] V ,
\end{equation}
is a vector correcting the dust-to-gas drift in the presence of coupling, and ${ \St = \tau \Omega }$ \!is \!the \!dust's Stokes number. Note that the total dust-to-gas drift is
\begin{equation}
    \label{eq:background_drift_total}
    \bv = \budo - \bugo = \bw + V \ey = \frac{- 2 \fgo \St }{1 + (\fgo \St)^{2}} \left[ \ex - \frac{\fgo \St}{2} \, \ey \right] V ,
\end{equation}
in agreement with \citetalias{YoudinGoodman05}. $V$ acts a characteristic velocity.

\vspace*{-0.5 \baselineskip}
\subsection{Linearised equations}
\label{sub:linearised_equations}

Let us now perturb the background flow, ${ \overline{f} }$, with a perturbation, ${ f' }$, which we assume to be small, ${ f' \ll \overline{f} }$. The total flow is then ${ f = \overline{f} + f' }$, where ${ f }$ denotes any variable (pressure, density, velocity). 

\newcommand{\hi}{h'} 
\newcommand{\rgi}{\varrho_{g}'} 
\newcommand{\bugi}{\ensuremath{\bb{u}_{g}'}} 
\newcommand{\rdi}{\varrho_{d}'} 
\newcommand{\budi}{\ensuremath{\bb{u}_{d}'}} 

\newcommand{\bC}{\mathbf{C}} 

At the linear order, this leaves the perturbation equations
\begin{subequations}
    \label{eq:perturbation_equations_SI}
    \begin{align}
         \grad \bcdot \bugi \, &= 0 , \label{eq:perturbation_equations_SI_continuity_gas} \\
         \!\! \left[ \pD t + \bugo \bcdot \grad + \bS \right] \bugi \, &= \! - \grad \hi \! + \! \bC \, \bugi \! +  \! \frac{\dtgo}{\tau} (\budi \! - \! \bugi + \! \bv \rdi) , \label{eq:perturbation_equations_SI_momentum_gas} \\
         \left[ \pD t + \budo \bcdot \grad \right] \rdi \, &= - \grad \bcdot \budi , \label{eq:perturbation_equations_SI_continuity_dust} \\
         \!\! \left[ \pD t + \budo \bcdot \grad + \bS \right] \budi \, &=  \bC \, \budi \! - \! \frac{1}{\tau} (\budi \! - \! \bugi) , \label{eq:perturbation_equations_SI_momentum_dust}
    \end{align}
\end{subequations}
where ${ \hi = P' / \rhog }$ is a pseudo-enthaply that conveniently replaces the pressure perturbation, ${ \rdi = \rho_{d}'/ \rhodo }$ is the relative perturbation in dust density, and ${\bC = 2 \Omega \, [ \ex \otimes \ey - \ey \otimes \ex ] }$ is the matrix representing the Coriolis operator ${ - 2 \Omega \, \ez \wedge \bcdot}$

Following \citetalias{YoudinGoodman05}, we shall only consider axisymmetric perturbations, \textit{i.e.} perturbations that do not depend on ${ y }$. As a consequence, the ${ \bS \bx \bcdot \grad }$ parts of the advective terms drop. To lighten notations, we also drop the overline ${ \overline{\phantom{a}} }$ over $\dtg$ from this point onward. It should not induce any ambiguity.

\thispagestyle{empty}
\enlargethispage{+1 \baselineskip}
\vspace*{-0.5 \baselineskip}
\section{Detailed physical picture of the SI}
\label{sec:core}

This section is the core of the paper, where we develop and justify the picture given in \S\ref{sec:intuition}.

We start in \S\ref{sub:dispersion_relation} by considering the dispersion relation, which shows an instability when the gas and dust are at resonance. The ensuing subsections explain the mechanism of this instability. First in \S\ref{sub:forward_action} we explain the forward action, by which an inertial wave concentrates dust. Then in \S\ref{sub:backward_reaction} we uncover the backward reaction, by which a dust density perturbation amplifies inertial waves. Finally in \S\ref{sub:combining_the_two_processes} we demonstrate how these two processes feed into each other at resonance, creating an unstable feedback loop -- the SI.

\subsection{Analysis of the dispersion relation}
\label{sub:dispersion_relation}

The goal of this first subsection is to extract as much information as possible from the dispersion relation. This will allow us to estimate the growth rate of the \RDI\ that is built on inertial waves, and to show that this instability and \SI\ are one and the same.

But first, let us establish the dispersion relation. To do this, we look for axisymmetric modal solutions to Eqs.~\eqref{eq:perturbation_equations_SI}, \textit{i.e.} solutions that can be decomposed into full Fourier modes of the form \smash{${ f(t, \bx) = \widehat{f} \, \e^{i (\bk \cdot \bx - \omega t)} }$} with ${ k_{y} = 0 }$. Such solutions exist if ${ \omega }$ and ${ k }$ satisfy
\begin{multline}
    \label{eq:dispersion_relation_SI}
    [\omega^{2} - (\hat{k}_{z} \kappa)^{2}] \, [\omega - \bv \cdot \bk] \, [\omega - \bv \cdot \bk + \frac{i}{\tau}] \, [(\omega - \bv \cdot \bk + \frac{i}{\tau})^{2} - \kappa^{2}] \\ 
    + \dtg \, P(\omega, \mu) = 0 ,
\end{multline}
where $P(\omega, \mu)$ is a polynomial in $\omega$ with no simple expression and ${ \kappa = \sqrt{2 \Omega (2 \Omega - S)} = \Omega }$ is the local epicyclic frequency.

Adimensionalising~\eqref{eq:dispersion_relation_SI} shows that the linear regime is controlled by four dimensionless numbers: the dust-to-gas ratio~$\dtg$, the Stokes number~$\St$, and the dimensionless wavenumbers~${ K_{x} = \eta r_{0} k_{x} }$ and~${ K_{z} = \eta r_{0} k_{z} }$.

\subsubsection{Resonance}
\label{ssub:resonance}

In the regime of test particles, ${ \dtg = 0 }$, the dispersion relation~\eqref{eq:dispersion_relation_SI} describes two gas inertial waves (${ \omega = \pm \hat{k}_{z} \kappa }$), a dust density wave (${ \omega = \bv \bcdot \bk }$), and three damped dust modes. Following \citetalias{SquireHopkins18a}, let us investigate what happens when a dust density wave and an inertial wave have similar frequencies.

\newcommand{\hh}{\widehat{h}} 
\newcommand{\hrg}{\widehat{\varrho}_{g}} 
\newcommand{\hbug}{\ensuremath{\bb{\widehat{u}}_{g}}} 
\newcommand{\hrd}{\widehat{\varrho}_{d}} 
\newcommand{\hbud}{\ensuremath{\bb{\widehat{u}}_{d}}} 

\newcommand{\bI}{\mathbf{I}} 

Test particles cannot exert a force on the gas, so the gas behaves as if it were isolated. Therefore, we can select the inertial wave of frequency ${ \omega = + \hat{k}_{z} \kappa }$ (where ${\hbk = \bk / k}$) and inspect its effect on the dust. We find
\begin{equation}
    \label{eq:explain_resonance_SI}
    \hrd = \frac{\mathbf{a} \bcdot \bk}{\hat{k}_{z} \kappa - \bv \bcdot \bk} \hh ,
\end{equation}
where $\mathbf{a}$ is a non-zero vector whose expression is of no interest here. Eq.~\eqref{eq:explain_resonance_SI} reveals that when ${ \hat{k}_{z} \kappa \approx \bv \bcdot \bk }$, the dust responds strongly to being forced by an inertial wave. This justifies the use of the term \lq resonance' in \lq resonant drag instability'.

Moreover, at perfect resonance, ${ \hat{k}_{z} \kappa = \bv \bcdot \bk }$, there is a singularity and the dust's response explodes. This indicates that the modal ansatz is inappropriate at resonance. We need to consider non-modal solutions to Eqs.~\eqref{eq:perturbation_equations_SI}, \textit{i.e.} solutions that are only Fourier decomposed in space, \smash{${ f(t, \bx) = \widetilde{f}(t) \, \e^{i \bk \cdot \bx} }$}. Note that we use a hat ${ \, \widehat{} \, }$ to denote the amplitude of a full Fourier mode, but a tilde ${ \, \widetilde{} \, }$ for the time-dependent amplitude of a spatial Fourier mode.

\thispagestyle{empty}
\enlargethispage{+1 \baselineskip}
\subsubsection{Instability}
\label{ssub:instability_from_dispersion_relation}

\newcommand{\omegaO}{\omega_{0}} 
\newcommand{\omegaI}{\omega_{1}} 

\newcommand{\tomegaO}{\breve{\omega}_{0}} 
\newcommand{\tkappa}{\breve{\kappa}} 

Let us now leave the regime of test particles and enter the regime of thin dust, ${ 0 < \dtg \ll 1 }$. 

\citetalias{SquireHopkins18b} showed that everything scales with $\dtg^{1/2}$, so we perform a Puiseux expansion of the form \smash{${ \omega = \omegaO + \dtg^{1/2} \, \omegaI + \dtg \, ... }$} with \smash{${ \omegaO \! = \! \hat{k}_{z} \kappa \! = \! \bv \bcdot \bk }$}. The tricky point is that $\bv$ depends on~$\dtg$, so it should be expanded as well. When $\St$, $K_{x}$ and $K_{y}$ are all contained between $\dtg^{1/2}$ and $\dtg^{-1/2}$, we get
\begin{equation}
    \label{eq:perturbed_eigenfrequency_from_dispersion_relation_SI}
    \omegaI^{2} = \frac{ 1 - \St^{2} }{1 + \St^{2}} \frac{ \omegaO^{2} }{2} - i \frac{ \tau \omegaO }{1 + \St^{2}} \frac{ \Omega^{2} + \omegaO^{2} }{2} ,
\end{equation}
which is equivalent to Eq.~32 of \citetalias{SquireHopkins18a}. Crucially, \smash{$\omegaI^{2}$} is not a positive real number. This proves that the back-reaction force from the dust onto the gas causes an instability: the inertial-wave \RDI.  We can estimate its growth rate
\begin{align}
    \gamma &= \sqrt{\dtg} \, \mathfrak{Im} (\omegaI) + \mathcal{O} (\dtg) , \nonumber \\
    &= \sqrt{\dtg} \, \Omega \, \frac{ \displaystyle \sqrt[4]{\left[ 1 - \St^{2} \right]^{2} \tomegaO^{4} + \left[ 1 + \tomegaO^{2} \right]^{2} \St^{2} \, \tomegaO^{2}}}{\displaystyle \sqrt{ 2 \, (1 + \St^{2}) }} \label{eq:growth_rate_from_dispersion_relation_SI} \\
    & \hspace{1.1 cm} \times \sin \left[ \frac{1}{2} \arctan \left( \frac{\St}{|1 - \St^{2}|} \frac{1 + \tomegaO^{2}}{\tomegaO} \right) \right] + \mathcal{O} (\dtg) , \nonumber
\end{align}
where ${ \tomegaO = \omegaO / \Omega }$ is the dimensionless resonant frequency. In the regime of small particles, ${ \St \ll 1 }$, Eq.~\eqref{eq:growth_rate_from_dispersion_relation_SI} simplifies to
\begin{equation}
    \label{eq:growth_rate_from_dispersion_relation_SI_small_St}
    \gamma = \frac{\sqrt{\dtg}}{2 \sqrt{2}} \, \St \left[1 + \hk_{z}^{2} \right] \Omega ,
\end{equation}
which is equivalent to Eq.~34 of \citetalias{SquireHopkins18a}.

\begin{figure}
    \centering
    \includegraphics[width = \linewidth]{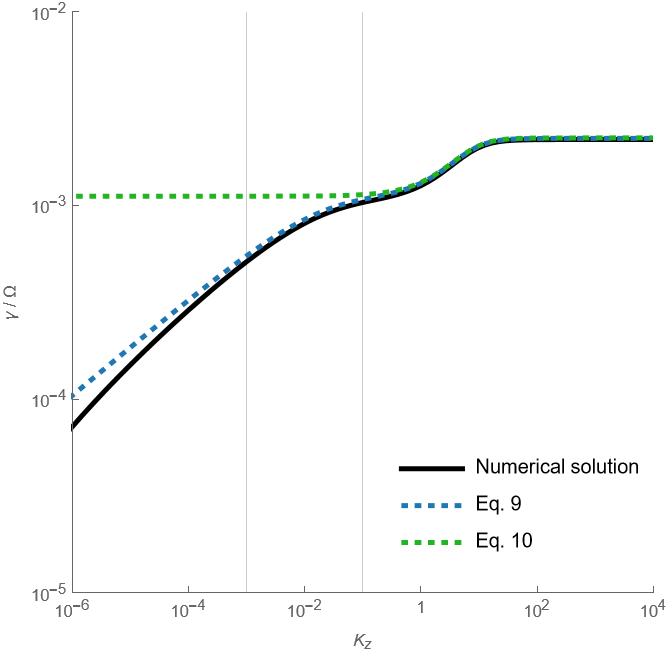}
    \caption{
        Dimensionless growth rate ${ \gamma / \Omega }$ of the inertial wave \RDI\ as a function of the vertical dimensionless wavenumber $K_{z}$. The solid black line displays the exact growth rate, as deduced from Eq.~\eqref{eq:dispersion_relation_SI}. For each value of $K_{z}$, $K_{x}$ was chosen so that the resonance condition ${ \hat{k}_{z} \kappa = \bv \bcdot \bk }$ would hold. As can be seen in Fig.~2 of \citetalias{SquireHopkins18a}, this means that $K_{x}$ scales with \smash{$K_{z}^{1/2}$} at low $K_{z}$ but becomes independent of $K_{z}$ at high $K_{z}$. The dashed blue line represents the prediction from Eq.~\eqref{eq:growth_rate_from_dispersion_relation_SI}. It seems valid in the regime of dilute dust, when ${ \dtg \ll \St, K_{x}, K_{z} }$. The dashed green line corresponds to the prediction from Eq.~\eqref{eq:growth_rate_from_dispersion_relation_SI_small_St}. It seems valid in the regime of dilute and well-coupled dust, when ${ \dtg, \St \ll K_{x}, K_{z} }$. Those limits are shown by the thin vertical lines. \\
        \textit{Parameters:} ${ \St = 10^{-1} }$, ${ \dtg = 10^{-3} }$.
    }
    \label{fig:growth_rate}
    \vspace*{-1 \baselineskip}
\end{figure}

This inertial-wave \RDI\ is the SI. Indeed, \citetalias{SquireHopkins18a} show in their Fig.~2 that the \RDI\ affects the same modes as the~\SI, and our Fig.~\ref{fig:growth_rate} shows that the growth rate of the \RDI\ is equal to the growth rate of the \SI.\footnote{The discrepancies at low $K_{z}$ in Fig.~\ref{fig:growth_rate} only appear because formulas~\eqref{eq:growth_rate_from_dispersion_relation_SI} and~\eqref{eq:growth_rate_from_dispersion_relation_SI_small_St} are valid when ${ K_{z} \gg \dtg^{1/2} }$.}

\enlargethispage{+1 \baselineskip}
We want to understand in detail the physics of this instability. The dispersion relation gave us the growth rate, and highlighted the role of a resonance between two waves. But we still need to understand how those two waves interact, and why it leads to an instability.

\subsection{Forward action: accumulation of dust by an inertial wave}
\label{sub:forward_action}

\RDIs\ arise from a feedback loop between two growth mechanisms. The first one, which we call the forward action, allows a gas wave to concentrate dust. 

For the acoustic \RDI, this process is straightforward: the perturbed velocities of a sound wave are convergent, and drag forces the dust to follow the gas, so dust concentrates (\citetalias{Magnan24}). For the \SI\ things are less clear, because inertial waves are incompressive: we need to understand how they can concentrate dust. We can best isolate this effect in the regime of test particles, ${ \dtg = 0 }$.

\subsubsection{There is an algebraic instability of dust forced by an inertial wave}
\label{ssub:forward_block_1}

In this limit, the gas is unaffected by the dust. So, the inertial wave should always exist, even at resonance. Simply, it will exert a drag force on the dust, which will react in a non-modal manner. Therefore, let us assume a modal inertial wave \smash{${ f_{g} (t, \bx) = \widehat{f_{g}} \, \e^{i (\bk \cdot \bx - \hk_{z} \kappa \, t)} }$} for the gas, and a non-modal answer \smash{${ f_{d} (t, \bx) = \widetilde{f_{d}}(t) \, \e^{i \bk \cdot \bx} }$} from the dust. Here \smash{$\widetilde{f_{d}}(t)$} is a time-dependent amplitude to be determined.

\newcommand{\trd}{\widetilde{\varrho}_{d}} 
\newcommand{\tbud}{\ensuremath{\bb{\widetilde{u}}_{d}}} 
\newcommand{\hugx}{\widehat{u}_{g, x}} 
\newcommand{\hugy}{\widehat{u}_{g, y}} 
\newcommand{\hugz}{\widehat{u}_{g, z}} 

The dust equations~(\ref{eq:perturbation_equations_SI_continuity_dust},~\ref{eq:perturbation_equations_SI_momentum_dust}) become
\begin{subequations}
    \label{eq:IVP_dust_SI}
    \begin{align}
        & \left[ \D t +  i (\bv \bcdot \bk) \right] \trd  = - i \, \bk \bcdot \tbud , \label{eq:IVP_dust_SI_continuity} \\
        & \left[ \D t + i (\bv \bcdot \bk) \bI + \bS \right] \tbud  = \bC \, \tbud - \frac{1}{\tau} (\tbud - \hbug \, \e^{- i \, \hk_{z} \kappa \, t}) . \label{eq:IVP_dust_SI_momentum}
    \end{align}
\end{subequations}
They can be combined to form the fourth-order \ODE
\begin{equation}
    \label{eq:forcing_in_kernel_of_operator_SI_dust}
    \mathcal{L}_{-1} \, \mathcal{L}_{0} \, \mathcal{L}_{+1} \left[ \D t + i (\bv \bcdot \bk) \right] \trd = \mathbf{b} \bcdot \hbug \, \e^{-i \, \hk_{z} \kappa \, t} ,
\end{equation}
where the ${ \mathcal{L}_{n} = \D t + i (\bv \bcdot \bk) + 1 / \tau + n \kappa }$ are operators linked to the damped dust modes of \S\ref{sub:dispersion_relation} and $\mathbf{b}$ is a vector whose exact expression is of no interest here. What matters is that $\mathbf{b}$ is not orthogonal to $\hbug$.

This form of the \ODE\ makes it clear that at resonance, the forcing frequency ${ \hk_{z} \kappa }$ collides with one of the dust's natural frequencies ${ \bv \bcdot \bk }$. Consequently, the particular solution to Eq.~\eqref{eq:forcing_in_kernel_of_operator_SI_dust} is proportional to ${ t \, \e^{-i \, \hk_{z} \kappa \, t} }$. This linear growth in time is the algebraic instability of \S\ref{sub:intuition_forward_action}.

\enlargethispage{-1 \baselineskip}
\subsubsection{It is the dust density perturbation that grows}
\label{ssub:forward_block_2}

To understand this algebraic instability, let us investigate what grows (and how fast) by solving Eqs.~\eqref{eq:IVP_dust_SI}.

Things are clearer in the reference frame that moves at the wave's phase speed ${ \mathbf{v}_{\varphi} \! = \! (\hk_{z} \kappa / k) \, \hbk }$. This adds an advection term and induces a Doppler shift, so Eqs.~\eqref{eq:IVP_dust_SI} become
\begin{align}
    & [ \D t +  i (\bv \bcdot \bk - \hk_{z} \kappa) ] \, \trd  = - i \, \bk \bcdot \tbud , \nonumber \\
    & [ \D t + i (\bv \bcdot \bk - \hk_{z} \kappa) + \bS ] \, \tbud  = \bC \, \tbud - \frac{1}{\tau} (\tbud - \hbug) . \nonumber
\end{align}
At resonance, this simplifies to
\begin{subequations}
    \label{eq:IVP_dust_SI_v2}
    \begin{align}
        & \D t \, \trd  = - i \, \bk \bcdot \tbud , \label{eq:IVP_dust_SI_continuity_v2} \\
        & [ \D t + \bS - \bC + \frac{1}{\tau} ] \, \tbud  = \frac{\hbug}{\tau} . \label{eq:IVP_dust_SI_momentum_v2}
    \end{align}
\end{subequations}
Note the structure of this system: $\trd$ is controlled by $\tbud$, which is itself controlled by $\hbug$. This means we can solve Eq.~\eqref{eq:IVP_dust_SI_momentum_v2} first. 

\newcommand{\budterm}{\bb{u}_{d}^{\text{ter}}} 

Its general solution is the sum of a homogeneous solution and a particular solution. The homogeneous solution is a linear combination of the eigenmodes of the dust velocity operator ${ \left[ \D t + \bS - \bC + 1 / \tau \right] }$. Its eigenfrequencies all have a negative real part, so the homogeneous solution quickly decays to zero. After a few dust stopping times $\tau$, all that is left is the particular solution, which is a constant. We denote it~$\budterm$ to emphasise that it can be interpreted as a terminal velocity.

Therefore, over long timescales, Eq.~\eqref{eq:IVP_dust_SI_continuity_v2} simplifies to
\begin{equation}
    \label{eq:algebraic_instability_forward_SI}
    \D t \, \trd  = - i \, \bk \bcdot \budterm ,
\end{equation} 
which reveals that the dust density perturbation grows linearly in time. Since it is the only dust feature to grow, we conclude that the forward action amplifies dust density waves. 

Equation~\eqref{eq:algebraic_instability_forward_SI} also gives the \lq algebraic growth rate' of the forward action,
\begin{equation}
    \label{eq:growth_rate_forward_SI}
    \gamma_{1} =  \frac{- i \, \bk \bcdot \budterm}{\trd (t = 0)} .
\end{equation}

\enlargethispage{-1 \baselineskip}
\subsubsection{Dust density perturbations grow via two mechanisms}
\label{ssub:forward_block_3}

${ i \, \bk \bcdot \budterm }$ translate to ${ \grad \bcdot \budterm }$ out of Fourier space. This tells us that the dust density grows because the secular dust motions are convergent. This is surprising, because drag makes dust follow the gas, and the gas motions are incompressive (and thus non-convergent).

To understand this, we need to calculate $\budterm$. It is defined by the equation ${ [\bI + \tau (\bS - \bC)] \, \budterm = \hbug }$, where $\bI$ is the identity matrix. This already shows that it is the shear $\bS$ and the Coriolis force $\bC$ that allow the dust velocity to differ from the gas velocity. Then by solving the equation, we find the radial and vertical components of $\budterm$,
\begin{subequations}
    \label{eq:components_terminal_dust_velocity_perturbation_SI}
    \begin{align}
    u_{d, x}^{\text{ter.}} &= \hat{u}_{g, x} + \,\, \frac{2 \, \St}{1 + \St^{2}} \, \hat{u}_{g, y} \, - \frac{\St^{2}}{1 + \St^{2}} \hat{u}_{g, x} , \label{eq:components_terminal_dust_velocity_perturbation_SI_x} \\
    u_{d, z}^{\text{ter.}} &= \hat{u}_{g, z} , \label{eq:components_terminal_dust_velocity_perturbation_SI_z}
    \end{align}
\end{subequations}
revealing that $\bS$ and $\bC$ cause a difference in \textit{radial} velocities.

This difference breaks the fine-tuned balance between $\hat{u}_{g, x}$ and $\hat{u}_{g, z}$ that made $\hbug$ orthogonal to $\bk$. This allows the dust motions to be compressive: 
\begin{align}
    -i \bk \bcdot \budterm &= -i \bk \bcdot \hbug - \frac{2 i \, \St \, k_{x}}{1 + \St^{2}} \hat{u}_{g, y} + \frac{i \, \St^{2} k_{x}}{1 + \St^{2}} \hat{u}_{g, x} , \nonumber \\
    &= \bigg( \frac{\tau \, k^{2}}{1 + \St^{2}} - i \frac{\tau \, \St \, \hk_{z} \, k^{2}}{1 + \St^{2}} \bigg) \, \hh . \label{eq:dust_compression}
\end{align}
So, we need to understand the physical mechanisms responsible for the two rightmost terms in Eq.~\eqref{eq:components_terminal_dust_velocity_perturbation_SI_x}.

The term proportional to $\hat{u}_{g, y}$ relies on a well-known mechanism. Remember the origin of the dust's radial drift: the gas is slightly sub-Keplerian and the dust Keplerian, so the dust feels a headwind, loses angular momentum, and moves inward. Well here, the inertial wave perturbs the gas' azimuthal velocity, so in some places the dust feels more (less) headwind than usual, loses more (less) angular momentum, and moves inward faster (slower), resulting in radial dust velocity perturbations. This explanation is supported by the similarity between the term of interest and Eq.~\eqref{eq:background_drift_total}. Fig.~\ref{fig:Forward_action_fast} shows that dust accumulates in pressure peaks of the inertial wave, and leaves pressure troughs. Indeed, the mechanism we just described is simply dust trapping by pressure bumps.

The other term's mechanism is more complex, so we leave its detailed description to \S\ref{sec:dust_grain_pusehd_radially}. To keep things here brief: the inertial wave perturbs the gas' radial velocity, creating radial winds that entrain the dust. However, an isolated particle does not change angular momentum when pushed radially,~so 
it cannot migrate radially, it just oscillates around its initial
orbit \!\!\! (\S\ref{sub:epicycles}). \!\!\! Of course, \!\!\! our \!\!\! dust \!\!\! particles \!\!\! are \!\!\! not \!\!\! isolated, \!\!\! they
interact with the gas via drag. Specifically, after a particle is pushed radially, it can exchange angular momentum with the background near-Keplerian gas flow. This allows dust particles to migrate radially and follow the gas' radial flows (\S\ref{sub:damped_epicycles}). But still, this angular momentum exchange process is not perfect, it cannot entirely erase the dust's radial stiffness, so the dust moves slower than the gas in the radial direction (\S\ref{sub:dust_grain_in_radial_wind}). Fig.~\ref{fig:Forward_action_slow} shows that this imperfect radial entrainment causes dust compression. Once again, our explanation is supported by the similarity between the term of interest and Eq.~\eqref{eq:difference_between_dust_velocity_and_wind}.

For small particles, the pressure-bump mechanism is much stronger than the radial wind mechanism, as it scales with $\St$ rather than $\St^{2}$. Consequently, the dust density perturbation grows almost exactly in phase with the pressure perturbation.

\subsubsection{Growth happens at resonance because the dust needs to not drift away from where it accumulates}
\label{ssub:forward_block_4}

Now that we understand the dust concentration mechanism, it is natural to wonder why it is only strong near resonance.

A clue comes from rewriting Eq.~\eqref{eq:algebraic_instability_forward_SI} outside resonance. It becomes
\begin{equation}
    \label{eq:algebraic_instability_forward_SI_v2}
    [ \D t +  i (\bv \bcdot \bk - \hk_{z} \kappa) ] \, \trd  = - i \, \bk \bcdot \budterm ,
\end{equation} 
which only exhibits an algebraic instability at perfect resonance. Otherwise, the advective term is non-zero, meaning that dust particles drift relative to the inertial wave. Precisely, what matters is that they drift out of waveplanes.

\!\!\! The role of resonance is to suppress this drift, so that particles cannot leave the plane they have been brought to by the mechanisms of \S\ref{ssub:forward_block_3}. This allows the dust density to increase forever in pressure bumps as more and more particles arrive. 

One may be concerned that this algebraic instability exists solely because we idealised the physics. Any detuning, however slight, kills the algebraic instability. So let us stress that the algebraic instability is not the \SI. The algebraic instability is just a fragile manifestation of one of the two robust processes behind the \SI: the process by which an inertial wave can cause large dust density perturbations.

Indeed, close to resonance, dust only drifts slowly, so the residence time of dust in pressure bumps is finite but long. This implies that a long transient regime of growth and a large final dust density is still possible at low detuning.

\subsection{Backward reaction: forcing of an inertial wave by the dust}
\label{sub:backward_reaction}

The preceding subsection focused on the first part of the \SI, in which an inertial wave causes a large dust density perturbation. Let us now see how a dust density perturbation can excite inertial waves. To observe this effect, we need the dust to have a significant effect on the gas. So we leave the regime of test particles and assume ${ \dtg > 0 }$.

\subsubsection{There is another algebraic instability}
\label{ssub:backward_block_1}

In any real system, the dust would be affected by the gas. But we can best isolate the process of interest in an idealised setup, in which the dust is decoupled from the gas. So let us assume a dust density wave ${ f_{d} (t, \bx) = \widehat{f_{d}} \, \e^{i (\bk \cdot \bx - \bv \cdot \bk \, t)} }$ that is not impacted by any gas perturbation. Remember that a dust density wave is a perturbation in dust density only, so the dust velocity remains unperturbed.\footnote{Because there is no restoring force, some may prefer to see this perturbation as an advected dust pattern rather than a wave.}

This dust density perturbation modifies the back-reaction force of the dust on the gas. Following our experience of \citetalias{Magnan24}, we expect the gas to respond in a non-modal way to this forcing. So let us assume a non-modal form for the gas properties, ${ f_{g} (t, \bx) = \widetilde{f_{g}}(t) \, \e^{i \bk \cdot \bx} }$.

We want to emphasise the effect of the dust density wave on the gas, so let us assume that the amplitude of the dust mode is much larger than that of the gas perturbation. This may only be true initially, if the gas perturbations gain amplitude over time. But while it is true, we can safely neglect the drag terms proportional to gas density or velocity in Eq.~\eqref{eq:perturbation_equations_SI_momentum_gas}, as well as an advection term.

\newcommand{\tH}{\widetilde{h}} 
\newcommand{\trg}{\widetilde{\varrho}_{g}} 
\newcommand{\tbug}{\ensuremath{\bb{\widetilde{u}}_{g}}} 

\vspace{+0.5 \baselineskip}
\noindent These assumptions reduce the gas equations~(\ref{eq:perturbation_equations_SI_continuity_gas}, \ref{eq:perturbation_equations_SI_momentum_gas}) to
\begin{subequations}
    \label{eq:IVP_gas_SI}
    \begin{align}
        & i \, \bk \bcdot \tbug = 0 , \label{eq:IVP_gas_SI_continuity} \\
        & [ \D t  + \bS] \, \tbug =  - i \, \bk \, \tH + \bC \tbug + \frac{\dtg}{\tau} \, \bv \, \hrd \, \e^{-i \, \bv \cdot \bk \, t} . \label{eq:IVP_gas_SI_momentum}
    \end{align}
\end{subequations}
The rightmost term summarises how the dust density perturbation affects the back-reaction. At equilibrium, the dust moving with velocity $\bv$ pushed the gas in front of it. Now that there is a perturbation, in regions where there is more dust than at equilibrium, the gas feels a stronger push than at equilibrium. Conversely, in regions where there is less dust than usual, the gas feels a weaker push. We will see in \S\ref{sub:combining_the_two_processes} that those equations emerge naturally when one studies coupled gas and dust.

\newcommand{\tugx}{\tilde{u}_{g, x}} 
\newcommand{\tugy}{\tilde{u}_{g, y}} 
\newcommand{\tugz}{\tilde{u}_{g, z}} 

With a bit of calculus, one can show that Eqs.~\eqref{eq:IVP_gas_SI} imply
\begin{equation}
    \label{eq:forcing_in_kernel_of_operator_gas_SI}
        \!\! \bigg[ \text{d}_{tt}^{2} + ( \hk_{z} \kappa )^{2} \bigg] \tugx = \hk_{z}^{2} \frac{\dtg}{\tau} \bigg[2 \Omega \, v_{y} - i (\bv \bcdot \bk) \,  v_{x} \bigg] \, \hrd \, \e^{-i \, \bv \cdot \bk \, t} .
\end{equation}
At resonance, the forcing frequency ${ \bv \bcdot \bk }$ coincides with the gas' natural frequency ${ \hk_{z} \kappa }$. Consequently, the particular solution to Eq.~\eqref{eq:forcing_in_kernel_of_operator_gas_SI} is proportional to ${ t \, \e^{-i \, \bv \cdot \bk \, t} }$, revealing that \smash{$\tugx$} grows linearly in time. So do \smash{$\tugy$}, \smash{$\tugz$} and \smash{$\tH$}. This is the second algebraic instability reported in \S\ref{sub:intuition_backward_reaction}.

\subsubsection{It is an inertial wave that grows}
\label{ssub:backward_block_2}

To find out what grows and why, let us solve Eqs.~\eqref{eq:IVP_gas_SI}. It will allow us to justify the heuristics of \S\ref{sub:intuition_backward_reaction}.

\vspace{+0.5 \baselineskip}
\noindent At resonance, we find
\begin{equation}
    \label{eq:solution_gas_over_time_SI}
    \!\!\!
    \begin{pmatrix}
        \tH (t) \vphantom{\displaystyle \frac{1}{1}} \\
        \! \tugx (t) \vphantom{\displaystyle \frac{1}{1}} \!\! \\
        \! \tugy (t) \vphantom{\displaystyle \frac{1}{1}} \!\! \\
        \! \tugz (t) \vphantom{\displaystyle \frac{1}{1}} \!\!
    \end{pmatrix}
    \! =
    \mathcal{A} \!
    \begin{pmatrix}
        \displaystyle \frac{\Omega}{k} \hk_{x} \vphantom{\displaystyle \frac{1}{1}} \!\! \\
        - \hk_{z} \vphantom{\displaystyle \frac{1}{1}} \!\! \\
        \displaystyle \frac{i}{2} \vphantom{\displaystyle \frac{1}{1}} \!\! \\
        \hk_{x} \vphantom{\displaystyle \frac{1}{1}} \!\!
    \end{pmatrix}
    \hrd \, t \, \e^{- i \, \hk_{z} \kappa \, t}
    \! + \text{complementary fun.}
\end{equation}
with \smash{${ \mathcal{A} = - \frac{\dtg}{2 \tau} ( \hk_{z} v_{x} + 2 i \, v_{y} ) }$} is a constant with the dimension of an acceleration. But the crucial information is that the vector on the right-hand side of Eq.~\eqref{eq:solution_gas_over_time_SI} is the eigenvector of an inertial wave. This confirms that a dust density wave can generate and amplify inertial waves.

Eq.~\eqref{eq:solution_gas_over_time_SI} also gives the `algebraic growth rate' of the backward reaction,
\begin{equation}
    \label{eq:growth_rate_backward_reaction}
    \gamma_{2} = - \frac{\dtg}{2 \tau} \bigg( \hk_{z} v_{x} + 2 i \, v_{y} \bigg) \times \frac{\Omega}{k} \hk_{x} \times \frac{\hrd}{\tH (t = 0)} .
\end{equation}

\subsubsection{It grows because the back-reaction force has the right pattern to accelerate the flows of an inertial wave}
\label{ssub:backward_block_3}

Equation~\eqref{eq:solution_gas_over_time_SI} indicates that if ${ v_{x} < 0 }$ but ${ v_{y} = 0 }$, then the pressure perturbation of the growing inertial wave is in phase opposition with the dust density perturbation. We draw this situation in Fig.~\ref{fig:Backward_reaction_fast}. The perturbed back-reaction drives radial gas flows, which appear to always be in the same direction as the inertial wave's radial velocity perturbations. So, the gas accelerates and the inertial wave gets stronger.

Conversely, if ${ v_{x} = 0 }$ and ${ v_{y} > 0 }$, then Eq.~\eqref{eq:solution_gas_over_time_SI} indicates that the growing inertial wave has its pressure pattern phase-shifted by ${ \varphi = \pi / 2 }$ with respect to the dust density pattern. We draw this situation in Fig.~\ref{fig:Backward_reaction_slow}. The same mechanism as above applies, except that this time the back-reaction accelerates the inertial wave's azimuthal flows.

In reality, Eqs.~\eqref{eq:background_SI} tells us that we have both ${ v_{x} < 0 }$ and ${ v_{y} > 0 }$, so the radial and azimuthal mechanisms co-exist. They negotiate to select an inertial wave whose phase shift is a compromise between $\pi/2$ and $\pi$, and allow it to grow. Now since ${v_{y} \ll |v_{x}| }$, the radial mechanism is stronger, and the growing inertial wave is almost exactly in phase opposition with the dust density perturbation.

\subsubsection{Growth happens at resonance because it needs the back-reaction force to always amplify the inertial wave}
\label{ssub:backward_block_4}

Now that we understand the growth mechanism, it is natural to ask why it is only active near resonance.

Essentially, a non-resonant wave sees its phase shift evolve over time. For instance, at a given instant, it may have a phase shift of ${ \varphi = \pi }$, in which case we are in the situation of Fig.~\ref{fig:Backward_reaction_fast}, and the wave gains amplitude. But a little while later, we have ${ \varphi = 0 }$ and the wave loses amplitude. These two situations happen equally often, hence why non-resonant waves do not grow. Only the resonant waves can maintain ${ \varphi }$ constant, remain at all times in the situation of Fig.~\ref{fig:Backward_reaction_fast}, and grow indefinitely.

That being said, the same point as in \S\ref{ssub:forward_block_4} stands: the backward reaction is fragile, but the process by which dust density waves amplify inertial waves is robust. This process may be transient outside resonance, but that is all \RDIs\ need.

\subsection{Combining the two processes to explain the RDI}
\label{sub:combining_the_two_processes}

The last two subsections described the two halves of the inertial-wave \RDI. In \S\ref{sub:forward_action} we saw how an inertial wave can concentrate dust, and in \S\ref{sub:backward_reaction} we saw how a dust density perturbation can amplify inertial waves. The present subsection offers to formalise our intuition that the \SI\ arises from a feedback loop between these two processes. To do so, we use an asymptotic method that calls upon the results of \S\ref{sub:forward_action} and~\ref{sub:backward_reaction}.

\subsubsection{Mathematical approach}
\label{ssub:combine_mathematical_framework}

Since we expect an exponential instability, let us fully Fourier-decompose each variable, ${ f(t, \bx) = \widehat{f} \, \e^{i (\bk \cdot \bx - \omega t)} }$. The perturbation equations~\eqref{eq:perturbation_equations_SI} become
\begin{subequations}
    \label{eq:Fourier-transformed_perturbation_equations_SI}
    \begin{align}
        & i \, \bk \bcdot \hbug = 0 , \label{eq:Fourier-transformed_perturbation_equations_SI_continuity_gas} \\
        & \begin{multlined}[t]
            \left[ - i \omega \, \bI - i \fd (\bw \bcdot \bk) \, \bI + \bS \right] \, \hbug = - i \bk \, \hh + \bC \hbug \\
            \hspace{+4.93 cm} + \frac{\dtg}{\tau} (\hbud - \hbug + \bv \, \hrd) ,
        \end{multlined} \hspace*{-2cm}\label{eq:Fourier-transformed_perturbation_equations_SI_momentum_gas} \\
        & \left[ -i \omega + i \fg (\bw \bcdot \bk) \right] \, \hrd = - i \, \bk \bcdot \hbud , \label{eq:Fourier-transformed_perturbation_equations_SI_continuity_dust} \\
        & \left[ -i \omega \, \bI + i \fg (\bw \bcdot \bk) \, \bI + \bS \right] \, \hbud = \bC \hbud - \frac{1}{\tau} (\hbud - \hbug) , \label{eq:Fourier-transformed_perturbation_equations_SI_momentum_dust}
    \end{align}
\end{subequations}

The dispersion relation given by Eq.~\eqref{eq:dispersion_relation_SI} is sufficient to show that there is an instability, and to estimate its growth rate~\eqref{eq:growth_rate_from_dispersion_relation_SI}. Unfortunately, it does not give us much insight into the mechanism of the instability.

\renewcommand{\omegaO}{\omega_{0}} 
\renewcommand{\omegaI}{\omega_{1}} 
\newcommand{\hhO}{\widehat{h}_{1}} 
\newcommand{\hhI}{\widehat{h}_{2}} 
\newcommand{\hbugO}{\widehat{\mathbf{u}}_{g, 1}} 
\newcommand{\hbugI}{\widehat{\mathbf{u}}_{g, 2}} 
\newcommand{\hbudO}{\widehat{\mathbf{u}}_{d, 1}} 
\newcommand{\hbudI}{\widehat{\mathbf{u}}_{d, 2}} 
\newcommand{\hrdO}{\widehat{\varrho}_{d, 0}} 
\newcommand{\hrdI}{\widehat{\varrho}_{d, 1}} 

To make progress on that front, we need to look closely at the eigenvectors and not just the eigenfrequencies. Our work on the acoustic \RDI\ (\citetalias{Magnan24}) and singular perturbation theory \citep{Seyranian03} suggests a solution of the form
\begin{alignat}{5}
    & \omega      &&= \omegaO      &&+ \sqrt{\dtg} \, \omegaI     &&+ \dtg \, \omega_{2} &&+ ... \, , \label{eq:Puiseux_expansion_SI} \\
    & \widehat{f} &&= \widehat{f}_{0} &&+ \sqrt{\dtg} \widehat{f}_{1}   &&+ \dtg \, \widehat{f}_{2} &&+ ... \, , \nonumber
\end{alignat}
where ${ \widehat{f} }$ represents any of ${ \hh, \hbug, \hrd}$ and $\hbud$. We also assume that \smash{${ \widehat{h}_{0} }$}, \smash{${ \widehat{\mathbf{u}}_{g, 0} }$} and \smash{${ \widehat{\mathbf{u}}_{d, 0} }$} are all equal to zero, but not ${ \omegaO }$ nor \smash{${ \hrdO }$}. This last assumption means that the dust density is more perturbed than any other variable.

Remember that there are other dimensionless parameters describing this system: the Stokes number ${ \St }$ and the dimensionless wave-numbers $K_{x}$, $K_{z}$. We shall make the same assumptions as in \S\ref{ssub:instability_from_dispersion_relation} and assume that they are all contained between $\dtg^{1/2}$ and $\dtg^{-1/2}$. This ensures that the sizes of all the terms in the perturbation equations~\eqref{eq:Fourier-transformed_perturbation_equations_SI} are dictated solely by the dust-to-gas ratio $\dtg$.

\vspace{-0.5 \baselineskip}
\subsubsection{Dust equations at order ${ \dtg^{0} }$}
\label{ssub:combine_dust_order_0}

At order ${ 1 }$, the dust equations (\ref{eq:Fourier-transformed_perturbation_equations_SI_continuity_dust}, \ref{eq:Fourier-transformed_perturbation_equations_SI_momentum_dust}) simply give
\begin{equation}
    \label{eq:order_0_dust_SI}
    \omegaO = \bv \bcdot \bk ,
\end{equation}
which shows that, to leading order, the mode frequency equals the dust advective frequency. This means our ansatz (in which the perturbation in dust density dominates) isolates the dust density wave. Our perturbative approach helps us study how this wave changes when we leave the test-particle limit.

\vspace{-0.5 \baselineskip}
\subsubsection{Gas equations at order ${ \dtg^{1/2} }$}
\label{ssub:combine_gas_order_1}

\newcommand{\hugOx}{\widehat{u}_{g, 1, x}} 
\newcommand{\hugOy}{\widehat{u}_{g, 1, y}} 
\newcommand{\hugOz}{\widehat{u}_{g, 1, z}} 

At order ${ \sqrt{\dtg} }$, the gas equations (\ref{eq:Fourier-transformed_perturbation_equations_SI_continuity_gas}, \ref{eq:Fourier-transformed_perturbation_equations_SI_momentum_gas}) give 
\begin{subequations}
    \label{eq:order_half_gas_SI}
    \begin{align}
         & i \, \bk \bcdot \hbugO = 0 , \label{eq:order_half_gas_SI_continuity} \\
         & \left[ i \omegaO \, \bI + \bS \right] \, \hbugO = -i \bk \, \hhO + \bC \, \hbugO . \label{eq:order_half_gas_SI_momentum}
    \end{align}
\end{subequations}
After some linear algebra, one can eliminate all the variables but pressure to get
\begin{subequations}
    \label{eq:order_half_gas_SI_sol}
    \begin{align}
        \omegaO &= \pm \hk_{z} \kappa , \label{eq:order_half_gas_SI_sol_omega} \\
        \hugOx &= \frac{- k_{z}}{\Omega \, \hk_{x}} \, \hhO , \,\,\,\,\,\, \hugOy = \frac{\pm ik}{2 \Omega \, \hk_{x}} \hhO , \,\,\,\,\,\, \hugOz = \frac{k}{\Omega} \, \hhO . \label{eq:order_half_gas_SI_sol_u}
     \end{align}
\end{subequations}
This is the frequency and structure of an inertial wave, so our leading-order gas perturbation must be an inertial wave. Note that only at resonance do Eqs.~\eqref{eq:order_half_gas_SI_sol_omega} and~\eqref{eq:order_0_dust_SI} agree. This emphasizes that our ansatz in powers of $\sqrt{\dtg}$ is only valid at resonance.

\subsubsection{Dust equations at order ${ \dtg^{1/2} }$}
\label{ssub:combine_dust_order_1}

It is at this order that things get interesting, because we recover the forward action of \S\ref{sub:forward_action}. Indeed, the dust equations~(\ref{eq:Fourier-transformed_perturbation_equations_SI_continuity_dust}, \ref{eq:Fourier-transformed_perturbation_equations_SI_momentum_dust}) give
\begin{subequations}
    \label{eq:order_half_dust_SI}
    \begin{align}
        -i \omegaI \, \hrdO &= -i \, \bk \bcdot \hbudO , \label{eq:order_half_dust_SI_continuity} \\
        \bS \, \hbudO &= \bC \, \hbudO - \frac{1}{\tau} (\hbudO - \hbugO) , \label{eq:order_half_dust_SI_momentum}
    \end{align}
\end{subequations}
which are identical, respectively, to Eq.~\eqref{eq:algebraic_instability_forward_SI} in \S\ref{ssub:forward_block_2} and the inline equation that defined $\budterm$ in \S\ref{ssub:forward_block_3}.

\!Solving Eq.~\eqref{eq:order_half_dust_SI_momentum} and feeding the result to Eq.~\eqref{eq:order_half_dust_SI_continuity} gives
\begin{equation}
    \label{eq:order_half_dust_SI_v2}
    -i \omegaI \, \hrdO = \bigg( \frac{\tau \, k^{2}}{1 + \St^{2}} -i \frac{\tau \, \St \, \hk_{z} \, k^{2}}{1 + \St^{2}} \bigg) \, \hhO ,
\end{equation}
which is analogous to the combination of Eqs.~\eqref{eq:algebraic_instability_forward_SI} and~\eqref{eq:dust_compression}. The difference is that here we assume exponential growth. Equation~\eqref{eq:order_half_dust_SI_v2} captures the two dust concentration mechanisms of \S\ref{sub:forward_action}. The key difference is that in \S\ref{sub:forward_action}, the forward action was isolated. Here it is connected to the backward reaction, and that connection is made clear through the neat asymptotic ordering.

\subsubsection{Gas equations at order ${ \dtg^{1} }$}
\label{ssub:combine_gas_order_2}

At order ${ \dtg }$, the gas equations~(\ref{eq:Fourier-transformed_perturbation_equations_SI_continuity_gas}, \ref{eq:Fourier-transformed_perturbation_equations_SI_momentum_gas}) become
\begin{subequations}
    \label{eq:order_one_gas_SI}
    \begin{align}
        & i \, \bk \bcdot \hbugI = 0 , \label{eq:order_one_gas_SI_continuity} \\
        & \!\!\! - \! i \omegaI \, \hbugO \! + \! \left[ -i \omegaO \, \bI \! + \! \bS \right] \hbugI \! = \! -i \bk \, \hhI + \bC \, \hbugI + \frac{\bv}{\tau} \hrdO , \label{eq:order_one_gas_SI_momentum}
    \end{align}
\end{subequations}
which are identical to Eqs.~\eqref{eq:IVP_gas_SI}, except that the time derivative $\D t$ becomes \smash{${ - i \omegaI \, \hbugO - i \omegaO \, \hbugI }$} due to our asymptotic ordering and our assumption of exponential growth. This similarity is a first hint that the dust density wave \smash{$\hrdO$} must amp-lify the inertial wave \smash{$ \{ \hhO, \hbugO \} $}. Furthermore, the last term of Eq.~\eqref{eq:order_one_gas_SI_momentum} shows that the dominant term in the back-reaction force comes from the dust density perturbation. This valida-tes the assumption we made in the third paragraph of~\S\ref{ssub:backward_block_1}.

\newcommand{\bev}{\mathbf{b}} 

One can eliminate pressure using the same trick as in \S\ref{ssub:backward_block_1}. One can also eliminate all the second-order terms by realising that the operator acting on \smash{$\hbugI$} is the operator that acted on \smash{$\hbugO$} in Eqs.~\eqref{eq:order_half_gas_SI}. We know from Eq.~\eqref{eq:order_half_gas_SI_sol_u} that this operator has a non-trivial kernel. Consequently, it cannot be surjective. And indeed, \smash{${ \bev = \ex + 2 i \hk_{z} \, \ey }$} is a vector that cannot be reached by the operator. Projecting the (pressure-free) momentum equation onto \smash{$\bev$} eventually leads to
\begin{equation}
    \label{eq:order_one_gas_SI_v2}
    -i \omegaI \!
    \begin{pmatrix}
        \hhO \vphantom{\displaystyle \frac{1}{1}} \\
        \hugOx \vphantom{\displaystyle \frac{1}{1}} \\
        \hugOy \vphantom{\displaystyle \frac{1}{1}} \\
        \hugOz \vphantom{\displaystyle \frac{1}{1}}
    \end{pmatrix}
    \! =
    \frac{\mathcal{A}}{\dtg} \!
    \begin{pmatrix}
        \displaystyle \frac{\Omega}{k} \hk_{x} \vphantom{\displaystyle \frac{1}{1}} \\
        - \hk_{z} \vphantom{\displaystyle \frac{1}{1}} \\
        \displaystyle \frac{i}{2} \vphantom{\displaystyle \frac{1}{1}} \\
        \hk_{x} \vphantom{\displaystyle \frac{1}{1}}
    \end{pmatrix} 
    \hrdO ,
\end{equation}
where $\mathcal{A}$ was defined in \S\ref{ssub:backward_block_2}. The tight similarity between Eqs.~\eqref{eq:solution_gas_over_time_SI} and~\eqref{eq:order_one_gas_SI_v2} confirms that the gas equations at order~$\dtg$ recover the backward reaction described in \S\ref{sub:backward_reaction}. But once again, this growth process was isolated in \S\ref{sub:backward_reaction}, whereas now it is connected to the forward action.

\subsubsection{Bringing it all together}
\label{ssub:combine_bringing_it_all_together}

Combining Eq.~\eqref{eq:order_half_dust_SI_v2} with the pressure component of Eq.~\eqref{eq:order_one_gas_SI_v2} gives
\begin{align}
    \omegaI^{2} &= \bigg( \frac{\tau \, k^{2}}{1 + \St^{2}} -i \frac{\tau \, \St \, \hk_{z} \, k^{2}}{1 + \St^{2}} \bigg) \times \frac{\Omega \, k_{x}}{2 \tau k^{2}} \bigg(\hk_{z} v_{x} + 2 i \, v_{y}  \bigg) , \nonumber \\
    &= - \gamma_{1} \gamma_{2} / \dtg \vphantom{\frac{1}{1}}, \nonumber \\
    & = \frac{ 1 - \St^{2} }{1 + \St^{2}} \frac{ \omegaO^{2} }{2} - i \frac{ \tau \omegaO }{1 + \St^{2}} \frac{ \Omega^{2} + \omegaO^{2} }{2} , \label{eq:growth_rate_squared_SI}
\end{align}
which is identical to Eq.~\eqref{eq:perturbed_eigenfrequency_from_dispersion_relation_SI}. Our asymptotic ordering framework is equivalent to the perturbation methods used in \S\ref{ssub:instability_from_dispersion_relation} or by Squire \& Hopkins. Admittedly, our algorithm is slower, but it provides deeper insights into the physics.

In particular, the second line of Eq.~\eqref{eq:growth_rate_squared_SI} shows that the growth rate of the \SI\ is deduced from the geometric mean of the \lq algebraic growth rates' of the forward action and the backward reaction. This shows nicely that the \SI\ is a feedback loop between these two growth processes. An inertial wave concentrates dust, and the ensuing dust density perturbation strengthens the inertial wave, which can then concentrate dust faster, etc...

\vspace{+0.5 \baselineskip}
\noindent It is important to recognise that \citetalias{SquireHopkins20} present the main ingredients of the instability mechanism, and specifically the idea that the \SI\ relies on a feedback loop between a forward and backward reaction. Our contribution is to better formalise this idea, thereby clarifying its complexities.

For instance, we now understand that the \SI\ fails in 2D (\textit{i.e.} when ${k_{z} = 0}$) because the resonance condition ${ \hk_{z} \kappa = \bv \bcdot \bk }$ only holds if ${ \bv = 0 }$, and that suppresses the backward reaction. As a side note, the radial-wind forward mechanism also fails, because there is no radial gas wind when ${ \bk \propto \ex }$.

\subsubsection{A point on phase-shifts}
\label{ssub:phase-shifts}

Remember that the forward action and the backward reaction both break into two mechanisms, one slow and one fast. The first term in the first parenthesis of Eq.~\eqref{eq:growth_rate_squared_SI} represents the slow forward mechanism, the second term in the same parenthesis represents the slow forward action, the first term  in the second parenthesis is the fast backward reaction, \textit{etc.}

The feedback loop between the two fast mechanisms is stable (it only causes a frequency correction). Indeed, multiplying the two fast terms in the first line of Eq.~\eqref{eq:growth_rate_squared_SI} gives a positive real number. The fast-fast loop fails because the pressure-bump mechanism concentrates dust in pressure maxima (Fig.~\ref{fig:Forward_action_fast}), whereas the radial backward reaction favours the inertial wave whose pressure maxima are in phase with dust density minima (Fig.~\ref{fig:Backward_reaction_fast}). Therefore, the~two mechanisms work against each other, they cannot cooperate.

As a result, it is the fast-slow and slow-fast feedback loops that drive the instability. Those loops can attain instability because the phase disagreement between their forward and backward mechanisms is lower, suggesting that the mechanisms are more amenable to cooperation. 

For instance, let us focus on the fast-slow loop. The pressure-bump mechanism favours a dust density in phase with pressure (Fig.~\ref{fig:Forward_action_fast}), whereas the azimuthal backward reaction favours the inertial wave whose pressure pattern is dephased by ${ \pi/2 }$ with respect to the dust density perturbation (Fig.~\ref{fig:Backward_reaction_slow}). Consequently, the two mechanisms can compromise by selecting the mode whose pressure is dephased by ${ \pi/4 }$ with respect to the dust density. On this middle ground, both mechanisms cooperate. 

To give this concept mathematical backing, let us artificially suppress the slow forward and fast backward mechanisms in Eqs.~\eqref{eq:order_half_dust_SI_v2} and~\eqref{eq:growth_rate_squared_SI}. First, the simplified Eq.~\eqref{eq:growth_rate_squared_SI} indicates that~$\omegaI$ has an argument of~$3 \pi / 4$. Then, the simplified Eq.~\eqref{eq:order_half_dust_SI_v2} indicates that \smash{${ \hhO \propto \e^{i \pi / 4 } \, \hrdO }$.}

\vspace{+0.5 \baselineskip}
\noindent Finally, we should acknowledge that we are not the first to realise that there are several possible feedback loops. \cite{Zhuravlev22} gives a brief description of the fast-fast and fast-slow loops at the end of subsection 3.4, and \cite{SquireHopkins20} note in section 4 that the fast-fast loop is stable. However, their explanation focuses on the slow-fast loop. This is unfortunate, because the fast-slow loop is the only one that can dominate (in the low-$\hk_{z}$ branch, where ${\omegaO \ll \Omega}$).

\vspace*{-1 \baselineskip}
\section{Discussion and conclusion}
\label{sec:conclusion}

This paper is the second of a pair investigating \RDIs. Those instabilities arise in mixtures of gas and dust, when the dust drifts at the right velocity to resonate with a gas wave. Our first paper studied the \RDI\ that involves sound waves (\citetalias{Magnan24}). The simplicity of this acoustic \RDI\ helped us gain an intuition for how all \RDIs\ develop.

The present paper focuses on an inertial-wave \RDI\ known as the \SI. This instability grows in \PPDs\ by tapping the free energy contained in the solids' radial drift (\citetalias{YoudinGoodman05}). Interestingly, the \SI\ saturates by forming dust filaments and clumps \citep{JohansenYoudin07}. If the clumps are dense enough, they can collapse gravitationally and form planetesimals \citep{Johansen+07}. This makes the \SI\ a promising way to overcome the metre-scale barrier of planet formation.

Unfortunately, the \SI\ is still poorly understood. This is not surprising, the coupled dynamics of gas and dust in a rotating disc are particularly complex. But it makes it hard to figure out how the instability saturates and to evaluate its robustness, thereby slowing down the community's efforts to assess the value of the \SI\ as a planetesimal formation theory.

Recognising this, several authors have already tried to explain the \SI. The earliest attempt was the `peloton' idea from \cite{YoudinJohansen07}. This mechanism may explain why dust clumps in the later stages of the instability, but it is non-linear so it cannot explain the onset of the instability. Interestingly, our work allows us to track down the issue: the peloton idea relies on the assumption that the main effect of a dust overdensity is to reduce the dust's drift velocity $\bv$, whereas we show in \S\ref{ssub:combine_gas_order_2} that it is actually to increase the back-reaction force of the dust on the gas. \cite{Jacquet+11} were the first to make a connection between the \SI\ and the well-known phenomenon of dust trapping in pressure maxima, and they also established the necessity of a Coriolis force to maintain those pressure bumps. However, their explanation is 2D whereas the \SI\ requires three dimensions. \cite{LinYoudin17} find a formal analogy between the dynamics of dusty gas and the thermodynamics of ideal gas, and use it to frame the \SI\ as a clockwise path in the ${ (P, V) }$ plane. While this is true, the analogy induces a layer of abstraction that impedes intuition. Finally, \citetalias{SquireHopkins20} propose an explanation founded on the powerful resonant drag paradigm. Our explanation ends up being similar to theirs, but delving into the mathematics allows us to clarify the picture and add details.

Like \citetalias{SquireHopkins20}, we posit that the \SI\ is due to a feedback loop. The `forward action' corresponds to the concentration of drifting dust by an inertial wave. It relies on two mechanisms. The fast one involves the tendency of a wave's pressure maxima to trap dust, the slow one is due to the dust's imperfect radial entrainment in the gas' radial winds (\S\ref{sec:dust_grain_pusehd_radially}). The `backward reaction' corresponds to the amplification of an inertial wave by a sinusoidal dust-density pattern. Essentially, variations in dust density perturb the drag force of the dust on the gas. Because this perturbed back-reaction force has a sinusoidal pattern, it excites an inertial wave in the gas. Since the dust's drift is mostly radial, the back-reaction force is mostly radial. But the dust's drift also has a small azimuthal component, so one can divide the backward reaction into two mechanisms as well: fast radial and slow azimuthal. It turns out that the fast-fast feedback loop is stable, so the \SI\ uses the fast-slow and slow-fast loops.\footnote{We think that another \RDI\ called the settling instability (\citetalias{SquireHopkins18a}) works in much the same way, except that the background dust-to-gas drift $\bv$ is vertical. This makes things much simpler, because the phase-shift issue that stabilises the fast-fast loop when $\bv$ is horizontal disappears. Therefore, to understand the onset of the settling instability, one only needs to understand one forward action, one backward reaction, and one feedback loop.}

We also explain the role of resonance. In the forward action, resonance ensures that dust cannot drift away from the pressure bump it has been brought to. In the backward reaction, it ensures that the inertial wave is always strengthened by the excess back-reaction from the dust density wave, and never weakened by phase mixing.

Our explanation is backed by rigorous mathematics. We formalise the feedback loop idea with a clear asymptotic theory that breaks up and clearly isolates the different components of the two loops, because they manifest at different orders. \S\ref{sub:combining_the_two_processes} offers a new way to compute the growth rate of \RDIs. Our procedure is slower than the linear algebra of Squire \& Hopkins, but more transparent. This makes it easier to understand the physics behind any given \RDI.

There are many physical effects that might be important in \PPDs\ and yet are neglected in the present study. We did not include self-gravity because we do not think it plays a key role at the linear stage (unless the disk is very massive). We only considered resonant modes because we learned in \citetalias{Magnan24} that \RDIs\ keep a high growth rate even when there are several percent of detuning. We did not include gas viscosity nor dust diffusion, mainly because their stabilising effect has already been well characterised \citep{Umurhan+20, ChenLin20}. Similarly, we assumed that all the dust particles have the same size even though \cite{Krapp19} and \cite{McNally21} showed that the \SI\ dwindles when the particle size distribution is broad. Finally and in no particular order, we did not include collisions between particles, we did not include any magneto-hydro-dynamic effect, we did not account for the dependence of the stopping time on the gas' velocity, and we did not account for any background turbulence.

We think that our asymptotic procedure will be a useful tool, helping to predict the impact of the aforementioned physical effects on the \SI. For instance, we used it in \citetalias{Magnan24} to study the effect of detuning and a varying stopping time.

\section*{Acknowledgments}

We wish to thank our referee (M. Pessah) as well as S.-J. Paardekooper, G. Laibe and V. V. Zhuravlev for their insightful comments. Support for N.M. was provided by a Cambridge International \& Isaac Newton Studentship.

\section*{Data Availability}

No new data were generated or analysed in support of this research.

\bibliographystyle{mnras}
\bibliography{main}

\appendix

\clearpage
\section{Dynamics of a dust grain pushed radially}
\label{sec:dust_grain_pusehd_radially}

To understand the slow forward mechanism, we need to figure out how dust grains respond to a radial gas flow. We shall do this progressively: first we will see how a particle responds to an instantaneous radial impulsion (\S\ref{sub:epicycles}), then we will add drag (\S\ref{sub:damped_epicycles}), and finally we will switch from an impulsion to a constant wind (\S\ref{sub:dust_grain_in_radial_wind}).

\subsection{Instantaneous radial impulsion without drag}
\label{sub:epicycles}

Let us start simple with a solid particle that is insensitive to drag, for instance because it is too big. We work in the (Keplerian) shearing box, so the particle moves according to
\begin{align}
    \ddot{x} &= + 2 \Omega \, \dot{y} + 3 \Omega^{2} \, x , \nonumber \\
    \ddot{y} &= - 2 \Omega \, \dot{x} , \nonumber
\end{align}
where ${x}$ is the particle's radial position, ${y}$ its azimuthal position, and ${ 3 \Omega^{2} x }$ the force deriving from the tidal potential~$\phi_{t}$. If we replace the azimuthal velocity $\dot{y}$ by the deviation to Keplerian velocity ${ \delta \dot{y} = \dot{y} + \frac{3}{2} \Omega x }$, the system simplifies to:
\begin{subequations}
    \label{eq:equations_of_motion_dust_particle_no_drag}
    \begin{align}
        \ddot{x} &= + 2 \Omega \, \delta \dot{y} , \label{eq:equations_of_motion_dust_particle_no_drag_x} \\
        \delta \ddot{y} &= - \textstyle \frac{1}{2} \Omega \, \dot{x} . \label{eq:equations_of_motion_dust_particle_no_drag_y}
    \end{align}
\end{subequations}

Assume that the particles starts at the origin with a small radial velocity. The initial condition are thus ${ \bx (0) = \mathbf{0} }$ and ${ \dot{\bx} (0) = u_{0, x} \, \ex }$, leading to
\begin{align}
    x (t) &= \,\,\,\, \textstyle \frac{u_{0, x}}{\Omega} \,\, \sin{(\Omega t)} , \nonumber \\
    y (t) &= 2 \, \textstyle \frac{u_{0, x}}{\Omega} \left[ \cos{(\Omega t)} - 1 \right] . \nonumber
\end{align}
Those equations indicate that the particle describes elliptic epicycles in the shearing box. Interestingly, ${ \delta \dot{y} = - \frac{1}{2} \Omega \, x }$, so the particle is sub-Keplerian when ${ x > 0 }$ and super-Keplerian when ${ x < 0 }$.

\begin{figure}
    \centering
    \includegraphics[width = 0.7 \linewidth]{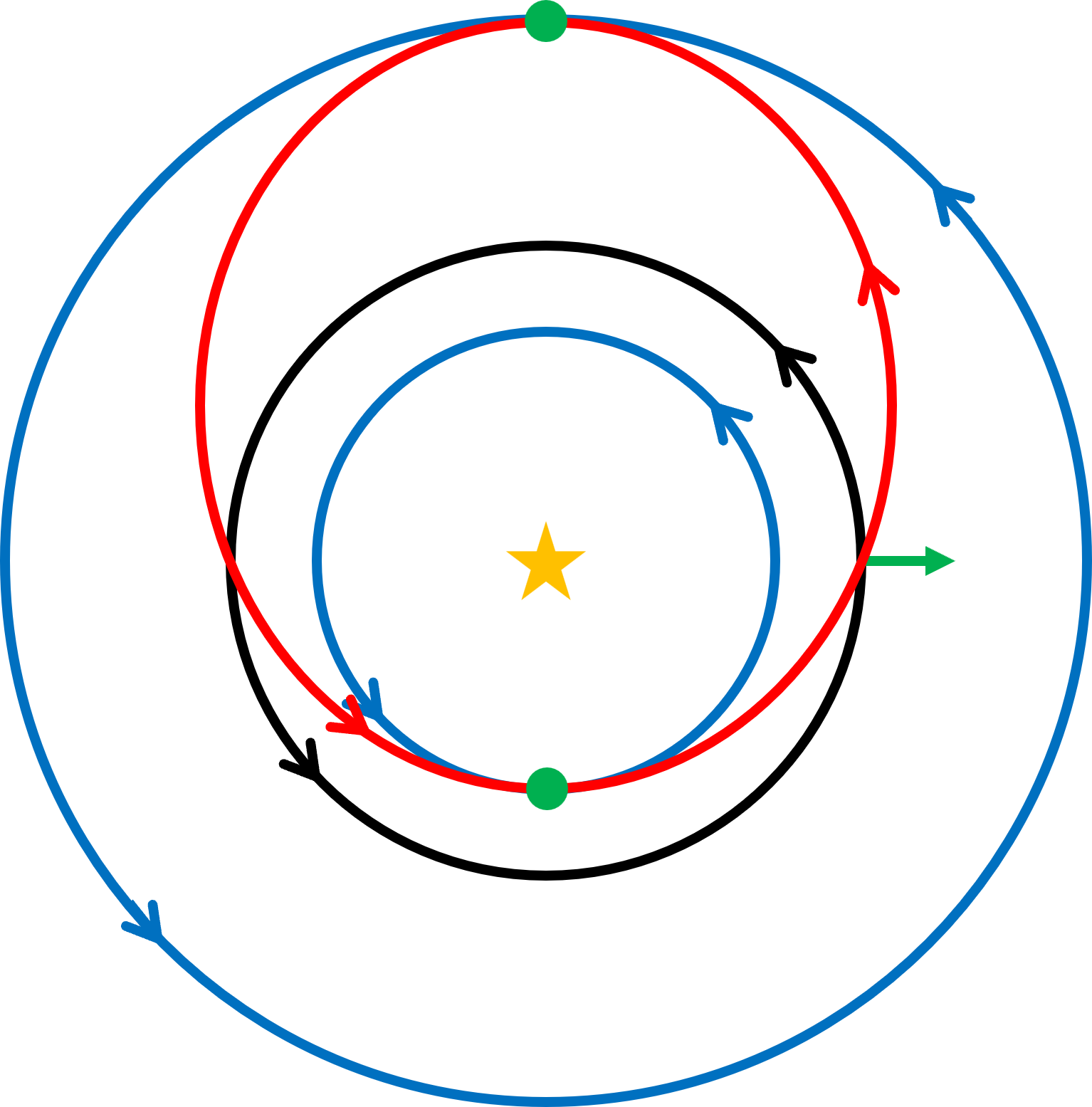}
    \caption{
        Orbital motion of a dust particle in the absence of drag. The particle starts on the black circular orbit and receives a small radial push indicated by the green arrow. It switches to the red elliptical orbit. The green dots represent apoapsis and periapsis. The specific angular momentum of the red orbit is the same as that of the black orbit, so the red velocity is lower than the blue velocity at apoapsis, and larger than the blue velocity at periapsis.
    }
    \label{fig:Motion_dust_particle_no_drag_global}
\end{figure}

Since there is no drag, this can all be reframed as a classical orbital mechanics question: ``A particle initially on a circular orbit receives an instantaneous radial push. How does it respond?". The answer is that the particle switches to the elliptical orbit that has the same specific angular momentum as the initial orbit. This process is depicted in Fig.~\ref{fig:Motion_dust_particle_no_drag_global}. 

Now remember that angular momentum is proportional to azimuthal velocity. Therefore, whenever the particle is outside of its initial orbit, its azimuthal velocity is lower than that of the local circular orbit. Conversely, whenever the particle is inside of its initial orbit, its azimuthal velocity is higher than that of the local circular orbit.

\subsection{Instantaneous radial impulsion with drag}
\label{sub:damped_epicycles}

Let us now perform the same calculation in the presence of drag. The findings of \S\ref{sub:epicycles} allow us to guess what will happen. 

Drag will damp any radial motion, so the particle will eventually settle on a circular orbit. It is not obvious which circle it will choose, because the particle feels a tailwind and gains angular momentum at apoapsis, but feels a headwind and loses angular momentum at periapsis. Fortunately, a theorem on the sign of alternating series allows us to affirm that the final radius will be larger than $r_{0}$ if the radial push was outward, and smaller than $r_{0}$ otherwise.

\begin{figure}
    \centering
    \includegraphics[width = \linewidth]{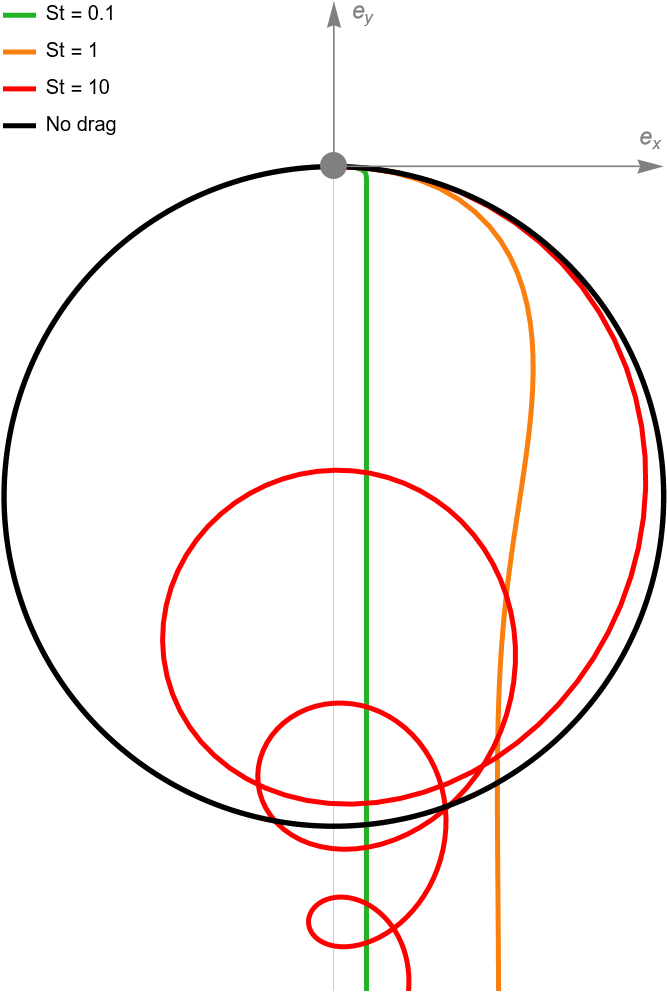}
    \caption{
        Trajectory of dust particles in the shearing box. All particles start at the centre of the box (grey dot), and receive a radial impulsion directed outward (to the right). The black particle does not experience drag, so it follows an epicycle. Note that the aspect ratio of the figure is ${1/2}$, so the ellipse appears circular. The green, orange and red particles experience different levels of drag. The marginally~coupled one, in orange, shows the largest secular radial displacement.
    }
    \label{fig:motion_dust_particle_drag_local}
\end{figure}

Indeed, once we include the drag exerted by Keplerian gas on the dust, the equations of motion~\eqref{eq:equations_of_motion_dust_particle_no_drag} become
\begin{subequations}
    \label{eq:equations_of_motion_dust_particle}
    \begin{align}
        \ddot{x} &= + 2 \Omega \, \delta \dot{y} - \frac{\dot{x}}{\tau} , \label{eq:equations_of_motion_dust_particle_x} \\
        \delta \ddot{y} &= - \textstyle \frac{1}{2} \Omega \, \dot{x} - \displaystyle \frac{\delta \dot{y}}{\tau} . \label{eq:equations_of_motion_dust_particle_y}
    \end{align}
\end{subequations}
Using the same initial conditions as in \S\ref{sub:epicycles}, we get
\begin{align}
    \! x (t) &= \! \frac{\tau \, u_{0, x}}{1 \! + \! \St^{2}} \! \left\{ \left[ 1 \! - \! \cos{(\Omega t)} \, \e^{- \frac{t}{\tau}} \right] \! + \! \St \sin{(\Omega t)} \, \e^{- \frac{t}{\tau}} \right\} , \nonumber \\
    \! y (t) &= 
    \begin{multlined}[t]
        \frac{\tau \, u_{0, x}}{1 \! + \! \St^{2}} \! \bigg\{ - \frac{3}{2} \Omega t \! + \! \frac{\St (1 \! - \! 2 \St^{2})}{1 \! + \! \St^{2}} \left[ 1 \! - \! \cos{(\Omega t)} \, \e^{- \frac{t}{\tau}} \right] \\
        + \! \frac{1 \! + \! 7 \St^{2}}{2 (1 \! + \! \St^{2})} \sin{(\Omega t)} \, \e^{- \frac{t}{\tau}} \bigg\} , \hspace{-0.5cm}
    \end{multlined} \nonumber
\end{align}
indicating damped epicycles, as shown in Fig.~\ref{fig:motion_dust_particle_drag_local}. If you wait long enough, the terms in ${\e^{-t/\tau}}$ disappear, leaving~only
\begin{align}
        x (t \gg \tau) &\approx \frac{\tau \, u_{0, x}}{1 + \St^{2}} , \nonumber \\
        y (t \gg \tau) &\approx \frac{\tau \, u_{0, x}}{1 + \St^{2}} \times - \frac{3}{2} \Omega t , \nonumber
\end{align}
meaning that the dust particle eventually settles on a circular orbit. The drag-induced circularisation of eccentric orbits was first calculated by \cite{Adachi+76}, but comparison is difficult because they frame the problem in terms of Keplerian orbital elements and include the dust's radial drift. Interestingly, dynamical friction induces a perfectly analogous circularisation \citep[see, e.g.,][]{PapaLala00,Bonetti20}   

As expected, the difference between the initial and final radii, ${ \Delta r = \frac{\tau \, u_{0, x}}{1 + \St^{2}} }$, takes the same sign as $u_{0, x}$. The biggest change in radius is obtained for marginally coupled particles. Indeed, small particles circularise almost immediately, so they do not have time to exchange much angular momentum with the gas. Conversely, big particles take several orbits to circularise, so they suffer from compensations between angular momentum gain at apoapsis and angular momentum loss at periapsis.

\subsection{Constant and uniform radial wind}
\label{sub:dust_grain_in_radial_wind}

Let us increase the complexity again by replacing the impulsion with a constant radial gas flow ${ \bug = u_{g, x} \, \ex }$.

After a few dust stopping times, the dust should reach a constant radial velocity. To estimate this velocity, we note that in \S\ref{sub:damped_epicycles}, the particle moved by roughly $\Delta r$ in a time $\tau$. Its mean radial velocity was therefore close to ${ \frac{\Delta r}{\tau} = \frac{u_{0, x}}{1 + \St^{2}} }$.

Formally, if the gas' azimuthal velocity is Keplerian, the radial wind transforms the equations of motion~\eqref{eq:equations_of_motion_dust_particle} to
\begin{subequations}
    \label{eq:equations_of_motion_dust_particle_with_wind}
    \begin{align}
        \ddot{x} &= + 2 \Omega \, \delta \dot{y} - \frac{\dot{x} - u_{g, x}}{\tau} , \label{eq:equations_of_motion_dust_particle_with_wind_x} \\
        \delta \ddot{y} &= - \textstyle \frac{1}{2} \Omega \, \dot{x} - \displaystyle \frac{\delta \dot{y}}{\tau} . \label{eq:equations_of_motion_dust_particle_with_wind_y}
    \end{align}
\end{subequations}
Those equations are identical to those that define $\budterm$ in \S\ref{ssub:forward_block_3}, provided the latter are written in components form, at resonance, and in the absence of any azimuthal gas velocity perturbation.


Whatever the initial conditions, after a few dust stopping times, we get ${ \dot{x}(t \gg \tau) \approx \frac{u_{g, x}}{1 + \St^{2}} }$, exactly as expected. From there, we deduce
\begin{equation}
    \label{eq:difference_between_dust_velocity_and_wind}
    \dot{x}(t \gg \tau) = u_{g, x} - \frac{\St^{2}}{1 + \St^{2}} u_{g, x} ,
\end{equation}
which is identical to Eq.~\eqref{eq:components_terminal_dust_velocity_perturbation_SI_x}, provided that there is no azimuthal gas velocity perturbation. This equation indicates that dust follows the radial wind, but is always slower than~it.

\subsection{Key takeaways}
\label{sub:dust_motion_in_wind_take_home_message}

In \S\ref{sub:epicycles} we learned that pushing a dust particle radially does not automatically change its long-term orbital radius. For instance, imagine an outward radial impulsion. Because it is radial, the particle does not gain any angular momentum. So it moves outward, but finds itself having a lower azimuthal velocity than the circular orbit at its new location. Consequently, it experiences a weak centrifugal force that cannot compensate gravity: the particle must fall back down.

Then in \S\ref{sub:damped_epicycles} we saw that if a dust particle is surrounded by gas in Keplerian rotation, then a radial push will lead to a long-term radial displacement. This is a consequence of drag, which allows the particle to exchange angular momentum with the gas. In our example, the dust particle becomes sub-Keplerian when it moves outward. Therefore, it feels a tailwind and gains angular momentum. Ultimately, this is what allows it to settle on a higher circular orbit.

Finally in \S\ref{sub:dust_grain_in_radial_wind} we found that dust can be entrained by a radial gas wind, but cannot keep up with that wind. This is due to the effect from \S\ref{sub:epicycles}: the weak centrifugal force does not compensate gravity, so there is a  net body force pointing towards the star. This inward force partially cancels out the outward drag force from the wind, hence why the dust moves slower than the gas.

\bsp	
\label{lastpage}
\end{document}